\documentclass[aps,pre,a4paper,twocolumn,groupedaddress, showpacs,showkeys,preprintnumbers,floatfix]{revtex4-1}

\usepackage[retainorgcmds]{IEEEtrantools}
\usepackage{graphicx}
\usepackage{psfrag}
\usepackage{color}
\usepackage{xcolor}
\usepackage{amsmath}
\usepackage{amscd}
\usepackage{amssymb}
\usepackage{amsthm}
\usepackage[section]{placeins}
\usepackage[normalem]{ulem}
\usepackage{caption}
\usepackage{subcaption}
\usepackage{soul}
\usepackage{nicefrac}
\usepackage{bbold}
\usepackage{cancel}
\usepackage{framed}
\usepackage{dsfont}
\usepackage{mathtools}
\usepackage{setspace}

\newcommand{\EX}{\mathds{E}}
\newcommand{\expect}[1]{\EX\left(#1\right)}
\newcommand{\expectc}[2]{\EX\left(#1 \,\middle\vert\, #2\right)}

\newcommand{\ppnc}{p_N^{\complement}}
\newcommand{\pprc}{p_R^{\complement}}
\newcommand{\psc}{S^{\complement}}
\newcommand{\ppsc}{S^{\complement}}
\newcommand{\ppoc}{P_0^{\complement}}

\begin{document}

\title{Emergence of echo chambers in a noisy adaptive voter model}
\date{\today}
\author{Andr\'e M. Timpanaro}
\email{a.timpanaro@ufabc.edu.br}
\affiliation{Universidade Federal do ABC,  09210-580 Santo Andr\'e, Brazil}
\begin{abstract}
Belief perseverance is the widely documented tendency of holding to a belief, even in the presence of contradicting evidence. In online environments, this tendency leads to heated arguments with users ``blocking'' each other. Introducing this element to opinion modelling in a social network, leads to an adaptive network where agents tend to connect preferentially to like-minded peers. In this work we study how this type of dynamics behaves in the voter model with the addition of a noise that makes agents change opinion at random. As the intensity of the noise and the propensity of users blocking each other is changed, we observe a transition between 2 phases. One in which there is only one community in the whole network and another where communities arise and in each of then there is a very clear majority opinion, mimicking the phenomenon of echo chambers. These results are obtained with simulations and with a mean-field theory.
\end{abstract}
\maketitle{}

\section{Introduction}

In recent years, the rise of social media as one of the main ways people interact with each other has raised concerns about their effects on the diversity of opinions in our society and on the popularization of extremist points of view \cite{echo-chamber-2, echo-chamber-3}. In this context, the emergence of echo chambers has been particularly worrisome \cite{echo-chamber-1, echo-chamber-4}. An echo chamber is a situation where a person is disproportionately exposed to points of view that align with their own, creating the illusion that their point of view is more common than it actually is and in some cases reinforcing these views due to confirmation biases.

In the case of social media, the ability of users to decide what content they consume (like deciding which other users to follow or block) might be an important ingredient in this phenomenon, if we assume that users prefer to be exposed to content aligned to their views. In this work we test this idea in a noisy adaptive voter model \cite{votante-def, votante-spontaneous-rewire, vot-adapt-1}, where besides the extra follow/block dynamic we consider a probability for agents to change their opinion at random. The main idea is that users will be connected in a symmetric way and connections between agents having different opinions may be rewired to become a connection between agreeing agents. We will implement the idea of following and blocking by connection rewiring and differentiate between 2 types of dynamic:

\begin{description}
    \item[Active rewires] When 2 agents interact, if their opinions are different, there is a probability that instead of the interaction taking place, their connection is rewired.
    \item[Reactive rewires] Whenever an agent changes opinion, there is a chance that its neighbours rewire their connections to it.
\end{description}

From a mathematical point of view, this means that we are using an adaptive network for our model \cite{livro-redes-adaptativas}. Adaptive networks are networks that change in accordance with the interactions between the agents inside them and they have garnered considerable attention in works studying the fragmentation of networks. In particular, there have been some works studying adaptive versions of the voter model (both using a copying rule and a majority rule) and in most of them a transition between complete fragmentation and consensus is found \cite{votante-spontaneous-rewire, vot-adapt-1, vot-adapt-2, vot-majoritario-adapt, vot-adapt-com-zealots, vot-adapt-noise, vot-adapt-external-field}. Our goal studying a noisy adaptive voter model was to see if the addition of noise would be sufficient to prevent fragmentation, which would be characterized by the formation of communities that are each close to a consensus. Ref \cite{vot-adapt-noise} did study a similar model, but no analysis of the network structure and its possible fragmentation was done and ref \cite{deffuant-adapt-noise} studied a noisy adaptive Deffuant model, observing community formation.
The study of the community structure is essential to understand if fragmentation is present or not, because the addition of the noise by itself is already enough to prevent the voter model from reaching consensus, even if the network is fixed. For this analysis we used the stochastic block model of community detection implemented in ref \cite{graph-tool}.

This work is organized as follows. In section \ref{sec:models} we define the active and reactive versions of our noisy adaptive voter model. In section \ref{sec:simulations} we present the simulation results, where we see a transition between a regime where the network has only one community and another where the different opinions fragment into different communities. In section \ref{sec:mf} a mean field theory is presented to try to explain the simulation results. Finally we summarize our conclusions in section \ref{sec:conc}

\section{Model Descriptions}
\label{sec:models}

We will be using a variant of the voter model \cite{votante-def}. In this model our social network is represented by a graph, where each of the nodes represent an agent and each of the edges represent a social connection between the corresponding agents. Each agent has an opinion, modeled by an integer, but there is no deeper meaning for these values besides mere labels. Self connections are not allowed, but we allow multiple connections between the same pair of agents (representing stronger connections), however for simplicity each of these connections is treated by the model as being different neighbours that happen to have the same opinion. Comparing with the usual voter model, we'll be adding 2 extra ingredients: The possibility of random opinion changes and the possibility of rewiring a connection between 2 agents, depending on their opinions. We discuss in this work two ways the rewiring can be implemented (active and reactive). The two versions will be studied separately and are described in the next sections.

Both models have 5 parameters: The number of sites $N$ and the average connectivity $q$, that parameterize the network (as it turns out, all other network details will be irrelevant); and the number of opinions $M$, the probability of a random opinion change (noise) $p_N$ and the probability of a rewiring happening $p_R$, that parameterize the dynamics between the agents. To avoid excessive repetition in the model descriptions, we define that:

\begin{itemize}
\item {\bf When we say an agent $i$ is affected by the noise}, it changes its opinion at random to one of the possible opinions ($1, \ldots, M$), chosen with equal probability.
\item {\bf When we say an agent $i$ attempts to rewire a connection to one of its neighbours $j$}, the following happens:
    \begin{enumerate}
        \item If $i$ is the only agent holding its opinion, then nothing happens.
        \item Otherwise, we remove one of the connections between $i$ and $j$ and we create a new connection between $i$ and some agent $k \neq i$, where $k$ is chosen at random uniformly among all the agents having the same opinion as $i$ (excluding $i$ itself).
    \end{enumerate}
\end{itemize}

\subsection{Active rewiring version}
\label{sec:active-rewires}

\begin{figure}[htbp!]
\centering
\includegraphics[width=0.45\textwidth]{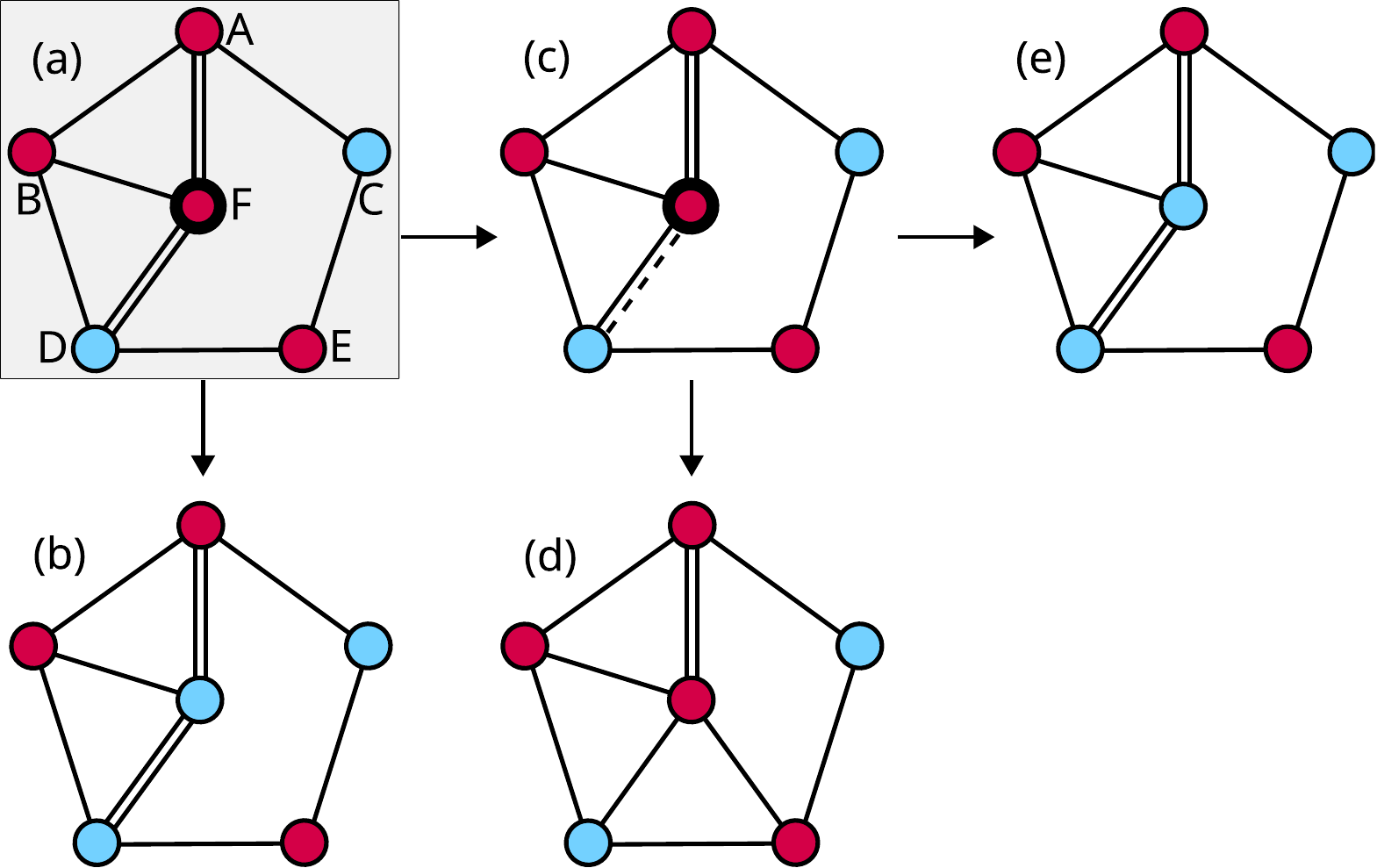}
\caption{Possible dynamics with active rewiring. {\bf (a)} Firstly an agent $F$ is chosen. {\bf (b)} $F$ might be affected by noise and change its opinion. {\bf (c)} Otherwise we choose one of $F$'s connections (the chosen connection is dashed) and $F$ interacts with the corresponding neighbour $D$ (note that since we draw a connection, more strongly connected neighbours have a higher change of being drawn). Finally, since $D$ and $F$ have different opinions, then either {\bf (d)} the chosen connection is rewired to another agent with the same opinion as $F$ ($E$ in this case) or {\bf (e)}  $F$ copies $D$'s opinion.}
\label{fig:ex-active}
\end{figure}

In this version, a rewiring can happen when 2 agents holding different opinions interact. The detailed time evolution is as follows (all possibilities are illustrated in figure \ref{fig:ex-active}):

\begin{enumerate}
\item At each time step, an agent $i$ is uniformly chosen at random (Fig \ref{fig:ex-active}a).
\item With probability $p_N$, $i$ is affected by the noise and we move on to the next time step (Fig \ref{fig:ex-active}b).
\item If $i$ wasn't affected by the noise and if $i$ has at least one neighbour, then we uniformly choose at random one of its connections. Let $j$ be the corresponding neighbour. If $i$ has no neighbours we move on to the next time step (Fig \ref{fig:ex-active}c). Note that this way of choosing the neighbour gives greater weight to neighbours that have multiple connections between them.
\item If $i$ and $j$ have the same opinion, nothing happens and we move on to the next time step.
\item However, if $i$ and $j$ have different opinions, then with probability $p_R$, agent $i$ tries to rewire its connection to $j$ (Fig \ref{fig:ex-active}d). Otherwise (probability $1-p_R$) we follow the usual voter model. That is, $i$ copies the opinion of agent $j$ (Fig \ref{fig:ex-active}e).
\end{enumerate}

\subsection{Reactive rewiring version}
\label{sec:reactive-rewires}

\begin{figure}[htbp!]
\centering
\includegraphics[width=0.45\textwidth]{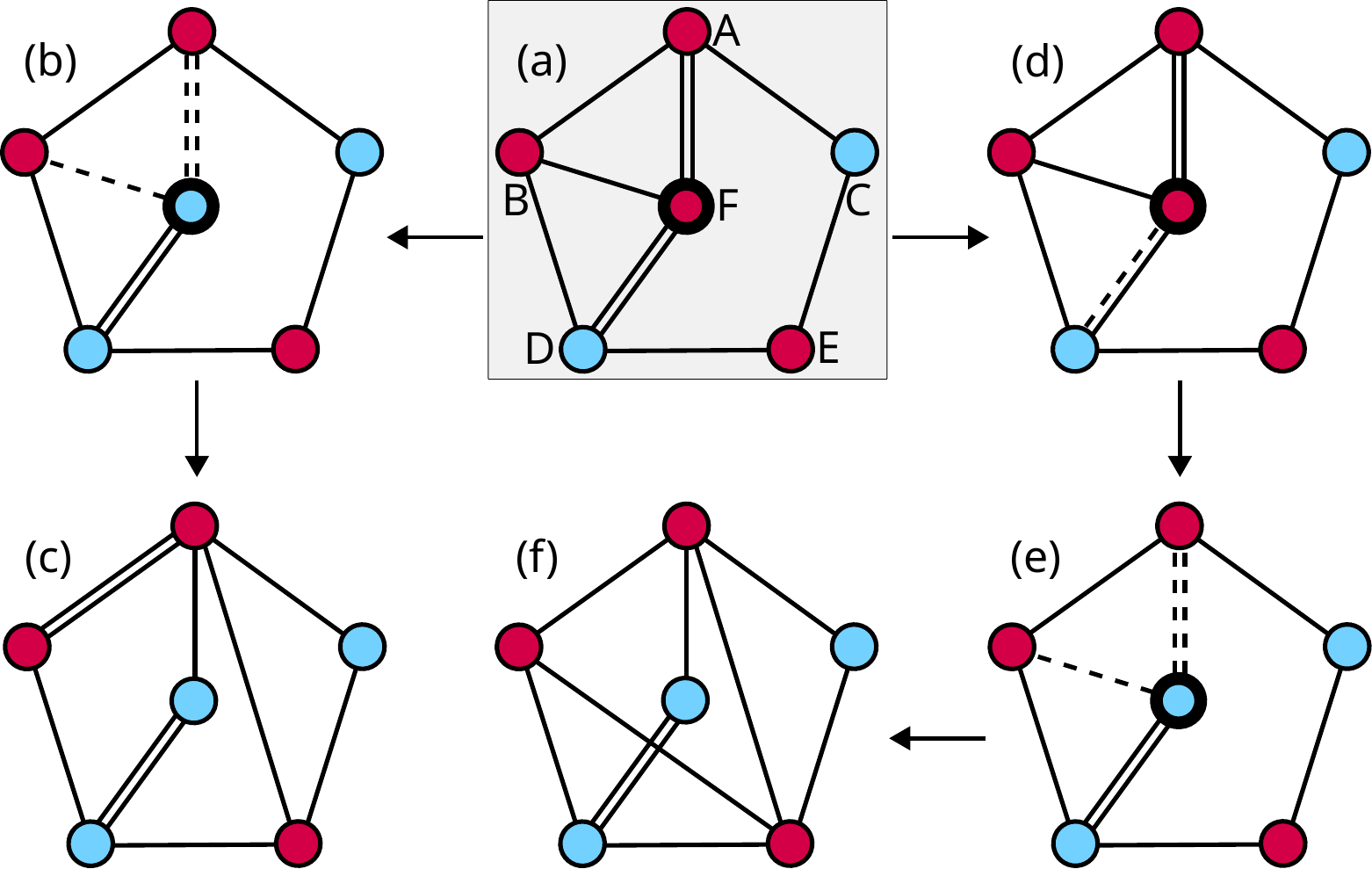}
\caption{Possible dynamics with reactive rewiring. {\bf (a)} Firstly an agent $F$ is chosen. {\bf (b)} $F$ might be affected by noise and change its opinion. $F$'s neighbours that do not share its new opinion will then attempt to rewire their connections with $F$ (the connections that may be rewired are dashed) {\bf (c)} In this case one of the connections with $A$ gets rewired to $E$, while the other remains intact and the connection with $B$ gets rewired to $A$. Note that in this case the neighbours are the ones doing the rewiring. {\bf (d)} If $F$ is not affected by the noise we choose one of $F$'s connections (the chosen connection is dashed). {\bf (e)} $F$ interacts with the corresponding neighbour $D$, copying $D$'s opinion. The opinion change once again causes $A$ and $B$ to attempt to rewire their connections with $F$. {\bf (f)} Here the connection with $B$ and one of the connections with $A$ gets rewired to $E$.}
\label{fig:ex-reactive}
\end{figure}

In this version, rewires can happen whenever an agent changes opinion. The detailed time evolution is as follows (all possibilities are illustrated in figure \ref{fig:ex-reactive}):

\noindent 

\begin{enumerate}
\item At each time step, an agent $i$ is uniformly chosen at random (Fig \ref{fig:ex-reactive}a).
\item With probability $p_N$, $i$ is affected by the noise. If $i$ retains the same opinion, nothing happens, but if this changes $i$'s opinion to $\sigma$ (Fig \ref{fig:ex-reactive}b), then each of its neighbours that have an opinion different from $\sigma$ attempt to rewire their connections to $i$ (if a neighbour has more than one connection, it attempts to rewire each one of them independently, so it's possible that they remain connected even if some rewires take place). We move on to the next time step (Fig \ref{fig:ex-reactive}c).
\item If $i$ wasn't affected by the noise and if $i$ has at least one neighbour, then we uniformly choose at random one of its connections. Let $j$ be the corresponding neighbour. If $i$ has no neighbours or if $i$ and $j$ have the same opinion, we move on to the next time step (Fig \ref{fig:ex-reactive}d).
\item We follow the usual voter model and $i$ copies the opinion of agent $j$. If this changes the opinion of $i$ to $\sigma$, then just like in step 2, each of its neighbours that have an opinion different from $\sigma$ attempt to rewire their connections to $i$ (Figs \ref{fig:ex-reactive}e and \ref{fig:ex-reactive}f).
\end{enumerate}

Note that if the opinion changes leave $i$ with its original opinion, then no rewires take place.

\section{Simulation Results}
\label{sec:simulations}

\begin{figure}[htbp!]
\centering
\includegraphics[width=0.45\textwidth]{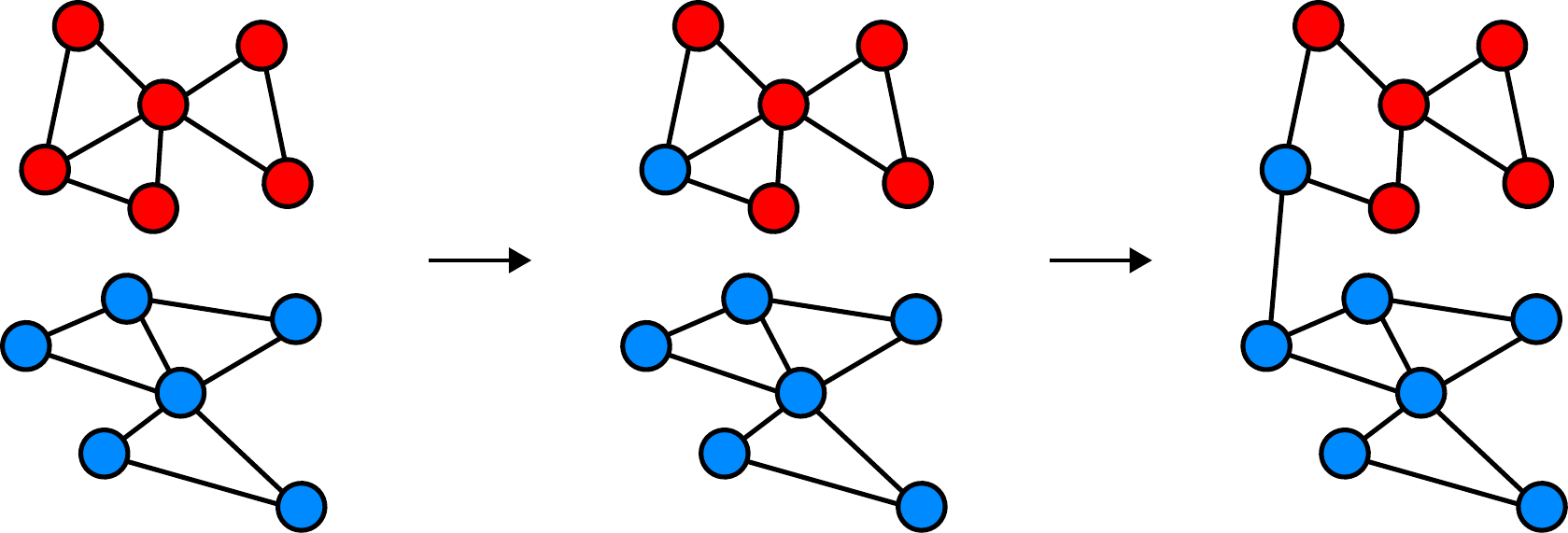}
\caption{The presence of noise and rewires can reconnect different components. Here we have an example using active rewires: In a given time step an agent changes its opinion from red to blue due to noise. In a later time step this same agent might rewire a connection after interacting with an agent holding opinion red. A similar process reintroduces connections when using reactive rewires.}
\label{fig:reconnect}
\end{figure}

The dynamics we are introducing creates a competition between 2 different mechanisms:

\begin{description}
\item[Rewiring] The main effect of the rewires is to move the network towards a fragmented state, where connections are highly assortative. As will be seen in the simulations, if this is the only mechanism present, then depending on the intensity of the rewiring the network may reach consensus before fragmenting or end up in a situation where each opinion has its own component.
\item[Noise] The noise prevents any community that arises in the network from being entirely composed of a single opinion. Furthermore, in conjunction with the rewiring dynamics, this means that any agent that changes to a new opinion inside of a community might end up being responsible for reintroducing connections with different communities (see figure \ref{fig:reconnect}). So the noise should act in the sense of keeping the communities from separating into different components of the network. Finally, the noise keeps the system in a state where each opinion is held by about the same number of agents (which is a feature already present in the usual voter model with noise).
\end{description}

\subsection{Irrelevance of the starting network}

\begin{figure}[htbp!]
\centering
\includegraphics[width=0.45\textwidth]{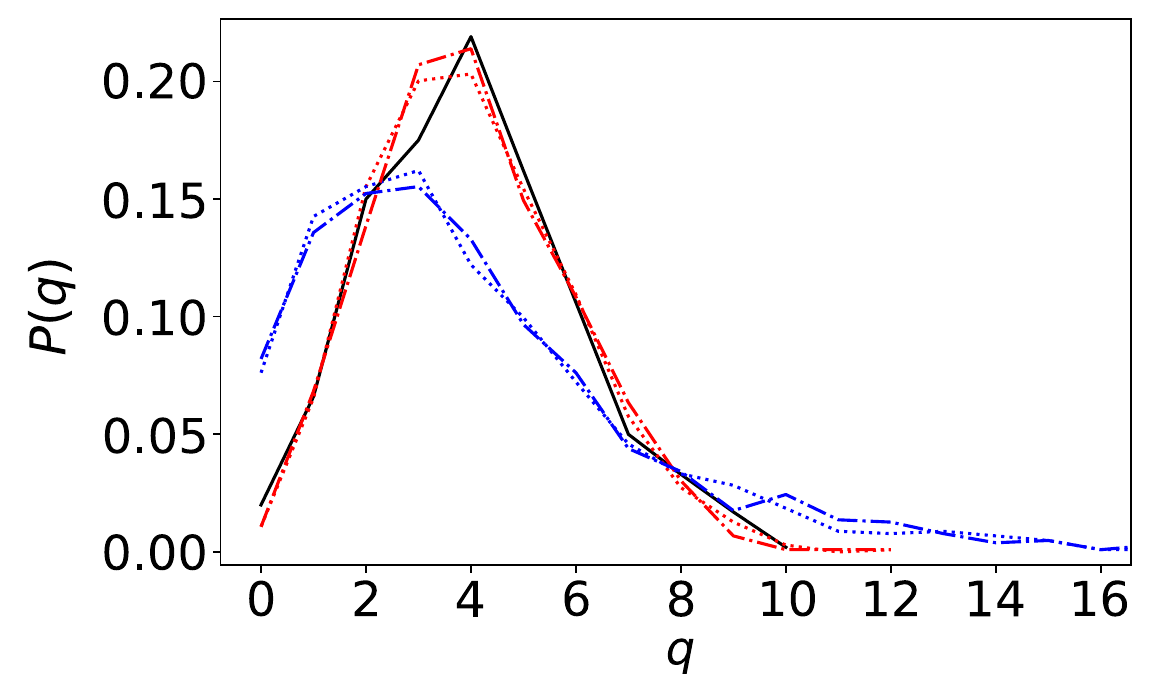}
\caption{Comparison between the degree distribution of an Erdös-Rényi graph \cite{erdos-renyi} and the final networks for different simulations. The full black line is the degree distribution for a realization of an Erdos-Rényi graph, the dotted lines are the final networks for simulations starting with a square network (with periodic conditions) and the dash-dotted lines are the final networks for simulations starting with a Barabási-Albert network \cite{rede-BA-def}. The red curves are simulations with active rewires and the blue ones are simulations with reactive rewires. In all cases the average coordination was 4 and the network size was 1024. The simulations also used $M = 4$, $p_N = 0.4$ and $p_R = 0.7$. The active rewire statistics seems to be well replicated by the Erdös-Rényi model (implying the connections are essentially random), while the reactive rewires lead to a slightly heavier tail.}
\label{fig:deg-dist}
\end{figure}

An important question when studying any opinion propagation model is the influence of the network topology in the dynamics. In our case, it turns out that the rewiring interaction destroys the original network structures and leads the network to a stationary state after a while. This is illustrated in figure \ref{fig:deg-dist} for the degree distributions

This means that the only aspects of the network that are relevant are the number of agents $N$ and the average coordination $q$, since these are invariant under the dynamics. Intuitively, this situation can be understood by noticing that after some time, all of the original edges will have been rewired, so the final network will reflect the fact that these new connections are being chosen according to the dynamics, which should have some stationary state.

\subsection{Community formation and detection}

\begin{figure}[htbp!]
\begin{subfigure}[t]{0.23\textwidth}
    \includegraphics[width=\textwidth]{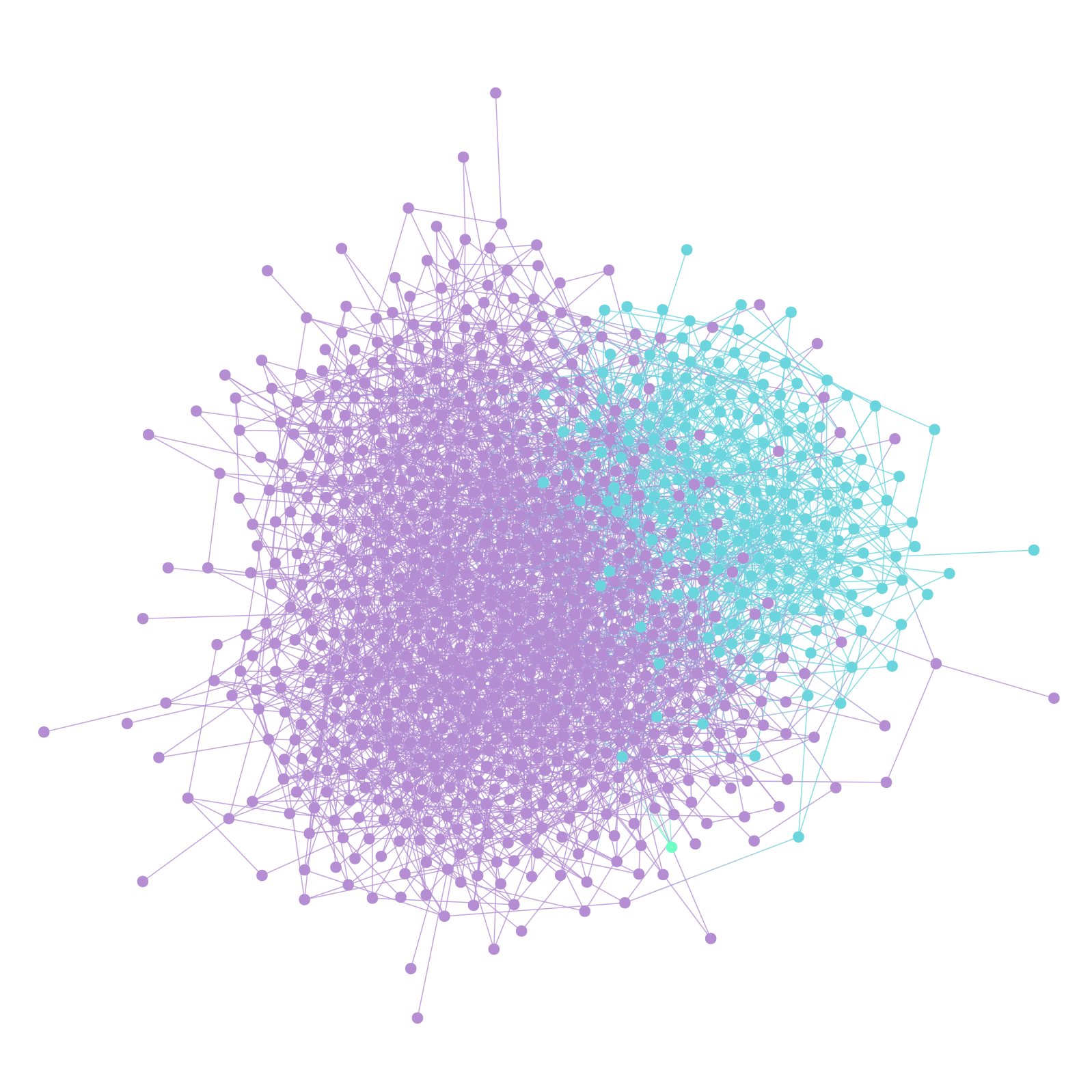}
    \captionsetup{justification=centering}
    \caption{Active. $q = 6$, $M=4$, $p_R=0.5$, $p_N=0.05$}
\end{subfigure}
\begin{subfigure}[t]{0.23\textwidth}
    \includegraphics[width=\textwidth]{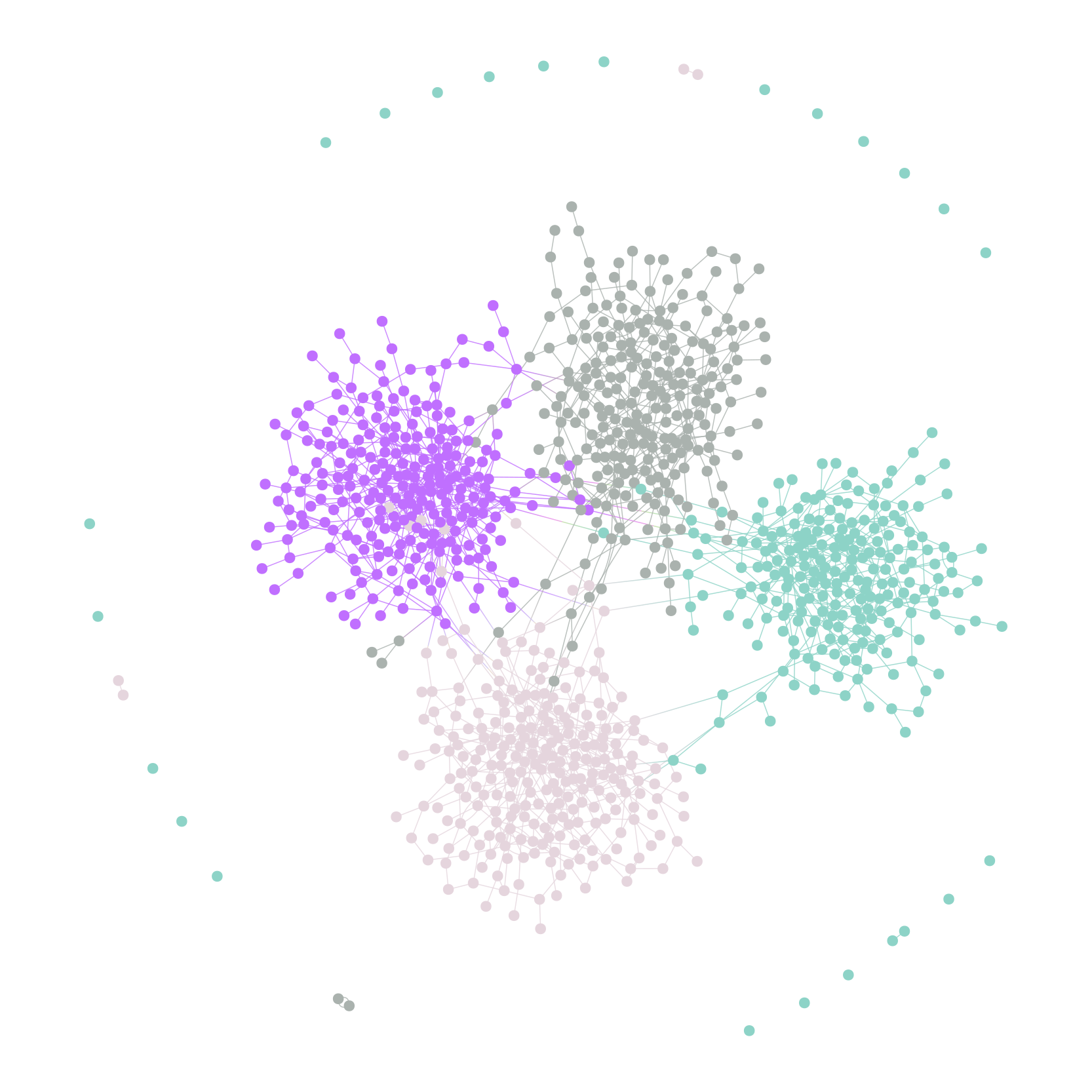}
    \captionsetup{justification=centering}
    \caption{Active. $q = 3$, $M=4$, $p_R=0.7$, $p_N=0.01$}
\end{subfigure}
\begin{subfigure}[t]{0.23\textwidth}
    \includegraphics[width=\textwidth]{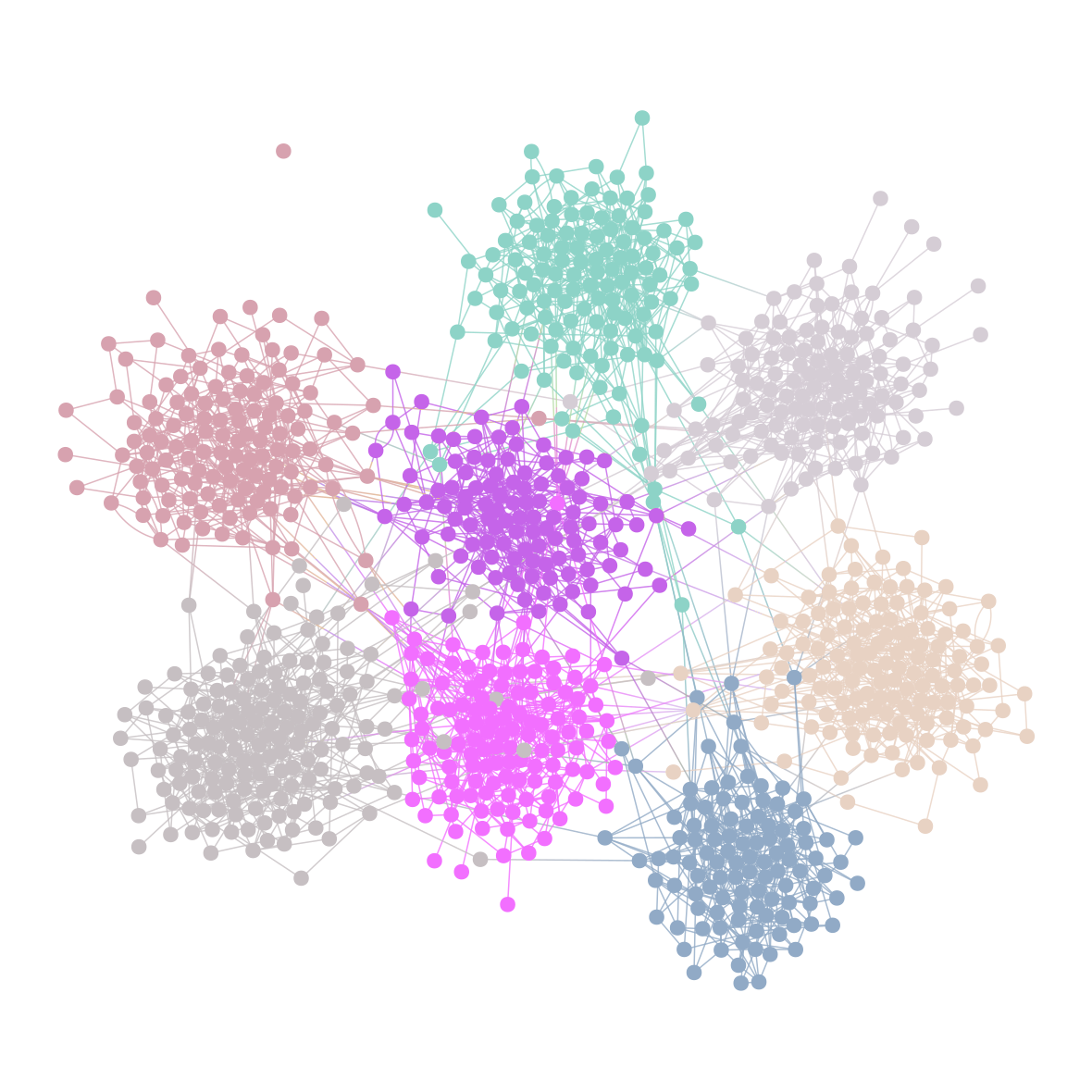}
    \captionsetup{justification=centering}
    \caption{Active. $q = 6$, $M=8$, $p_R=0.8$, $p_N=0.01$}
\end{subfigure}
\begin{subfigure}[t]{0.23\textwidth}
    \includegraphics[width=\textwidth]{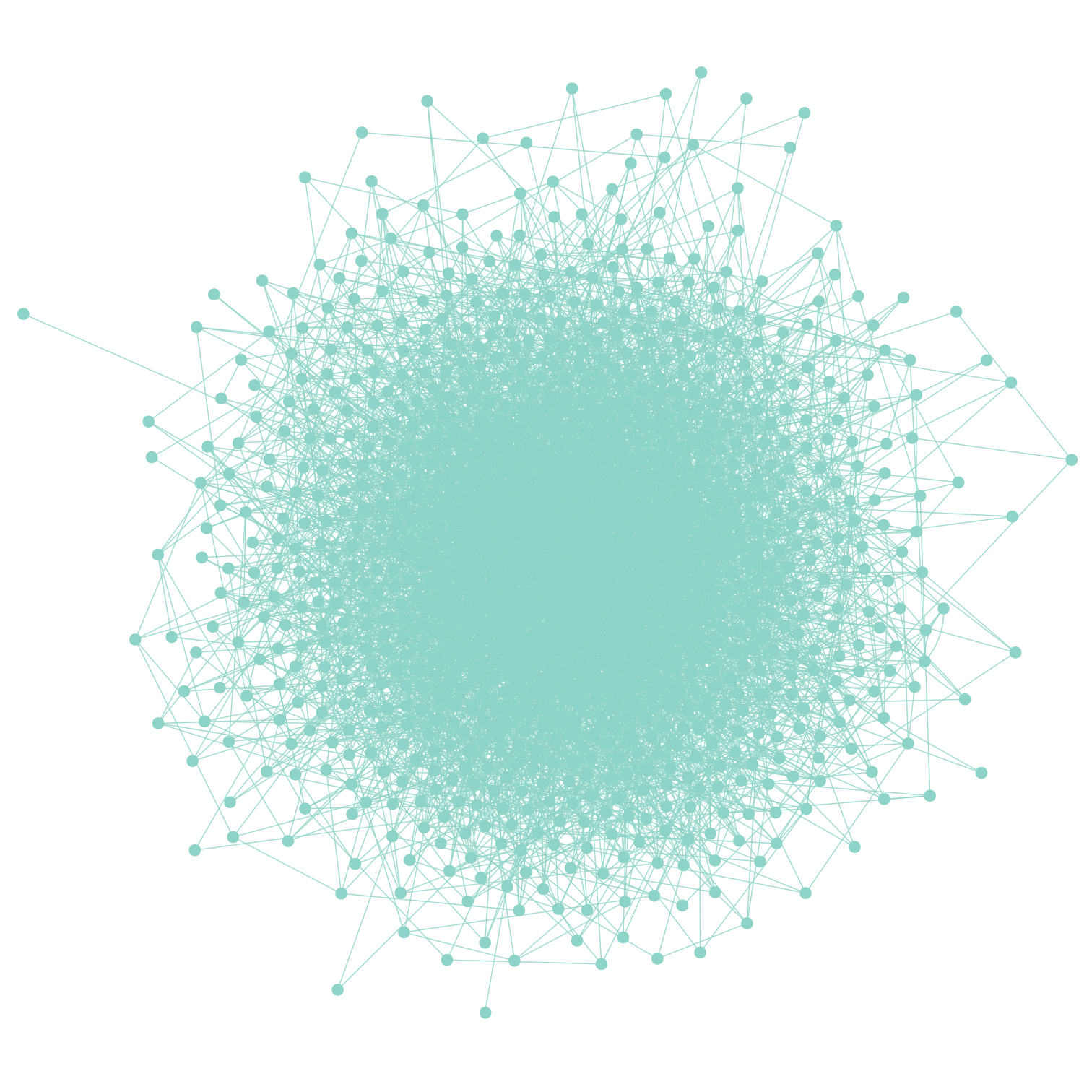}
    \captionsetup{justification=centering}
    \caption{Reactive. $q = 3$, $M=4$, $p_R=0.25$, $p_N=0.01$}
\end{subfigure}
\begin{subfigure}[t]{0.23\textwidth}
    \includegraphics[width=\textwidth]{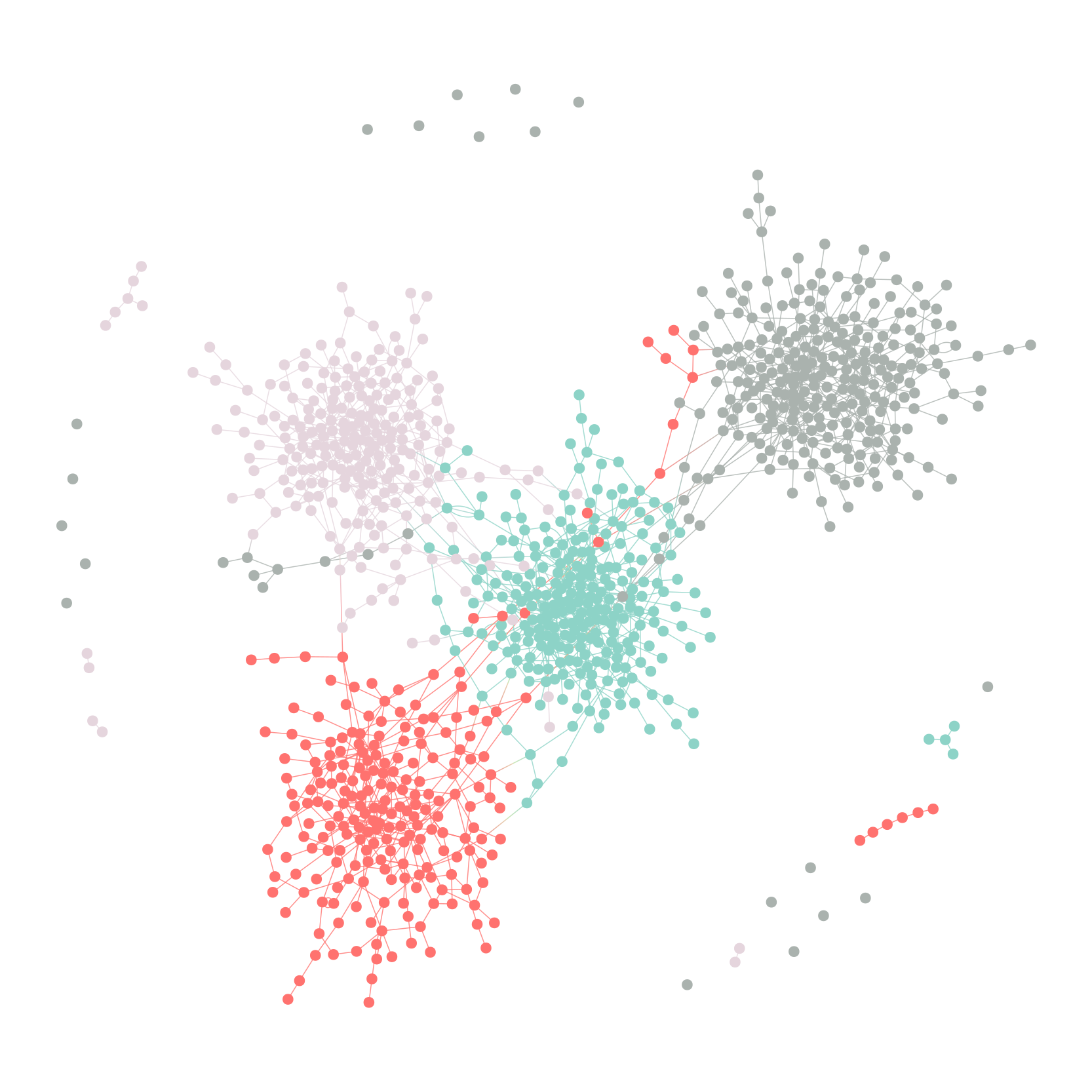}
    \captionsetup{justification=centering}
    \caption{Reactive. $q = 3$, $M=4$, $p_R=0.3$, $p_N=0.01$}
\end{subfigure}
\begin{subfigure}[t]{0.23\textwidth}
    \includegraphics[width=\textwidth]{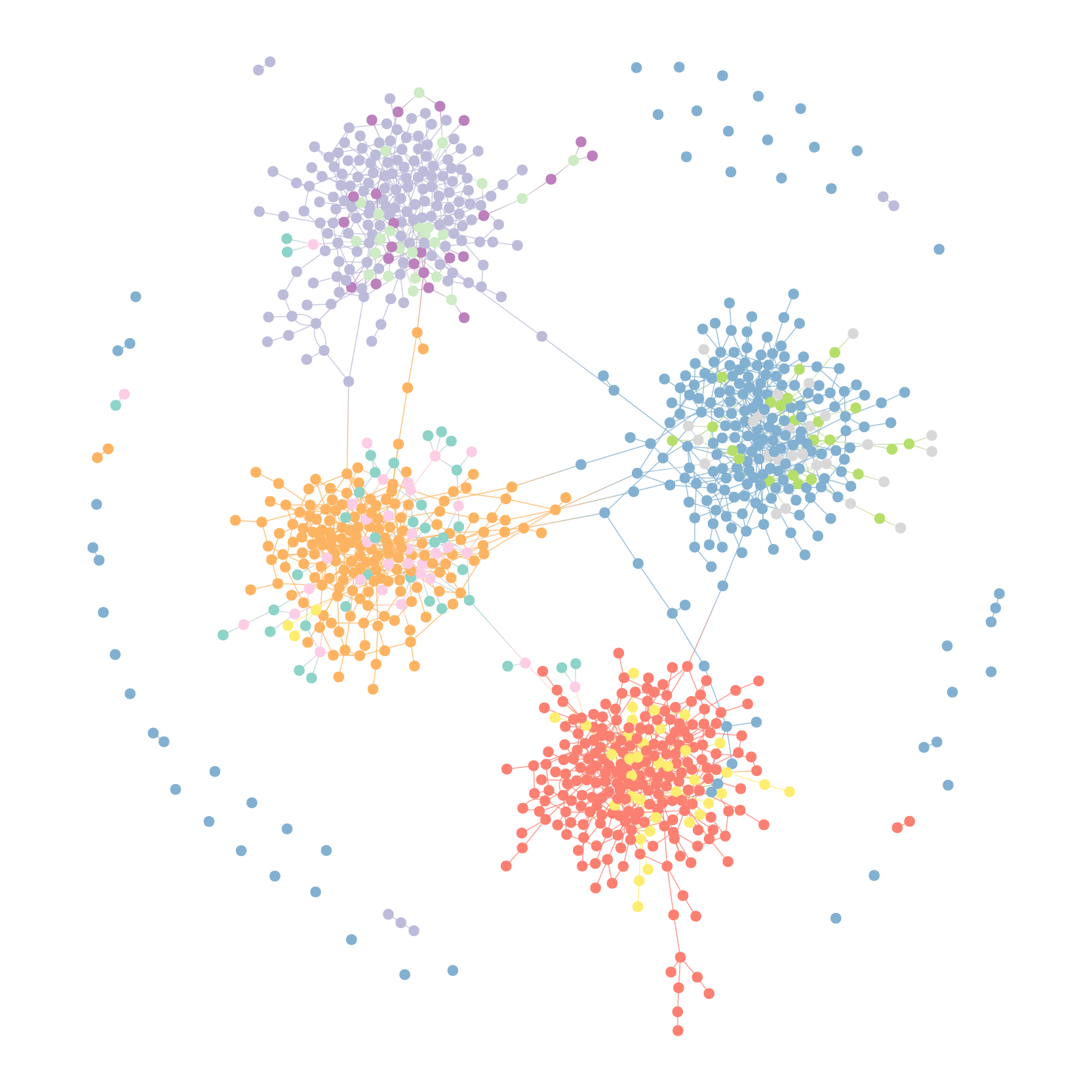}
    \captionsetup{justification=centering}
    \caption{Reactive. $q = 3$, $M=4$, $p_R=0.4$, $p_N=0.01$}
\end{subfigure}
\caption{Snapshots of different simulations in the presence of noise, showing the possible behaviours as well as some artifacts that can happen with SBM detection (all simulation use $N=10^3$). In (a) and (d) we have situations where there was no community formation, however the SBM run we chose incorrectly detects a second community in (a). This type of community splitting is an artifact of the algorithm that needs to be dealt with in the analysis and can also be seen in a smaller degree in (f). Comparing (b) and (c) we can see that in the regime where communities are formed, we must have $M$ communities due to symmetry (since the noise leads all opinions to show up in about the same proportion). Finally, the sequence (d-f) shows how abrupt community formation can be in the case of reactive rewires, as $p_R$ is increased. Also, comparing the active and reactive cases, we can see a trend where a smaller $p_R$ is enough for communities to form in the reactive case.}
\label{fig:communities}
\end{figure}

\begin{figure}[htbp!]
\centering
\begin{subfigure}[t]{0.3\textwidth}
    \centering
    \includegraphics[width=0.8\textwidth]{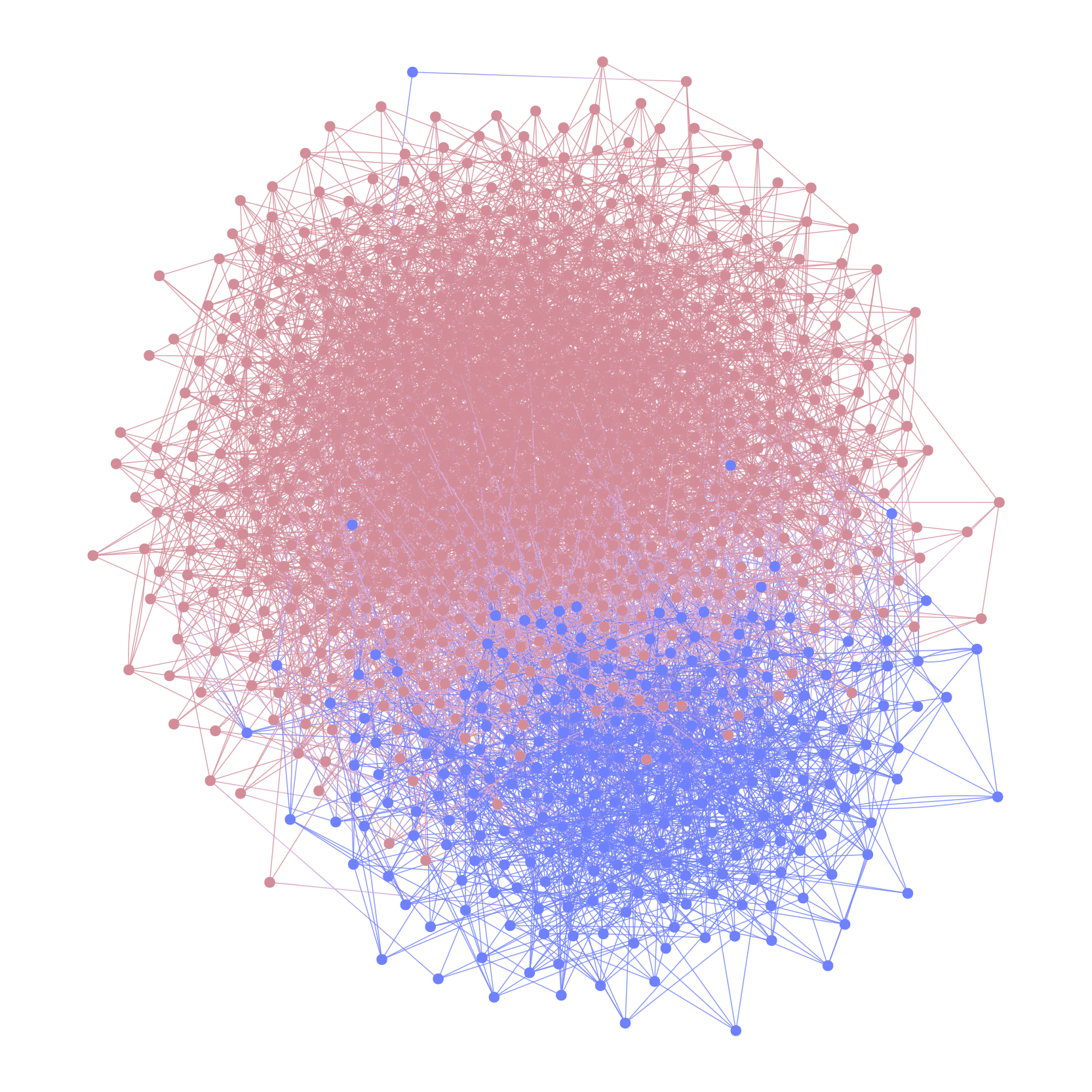}
    \caption{$p_R=0.6$}
\end{subfigure}
\begin{subfigure}[t]{0.3\textwidth}
    \centering
    \includegraphics[width=0.8\textwidth]{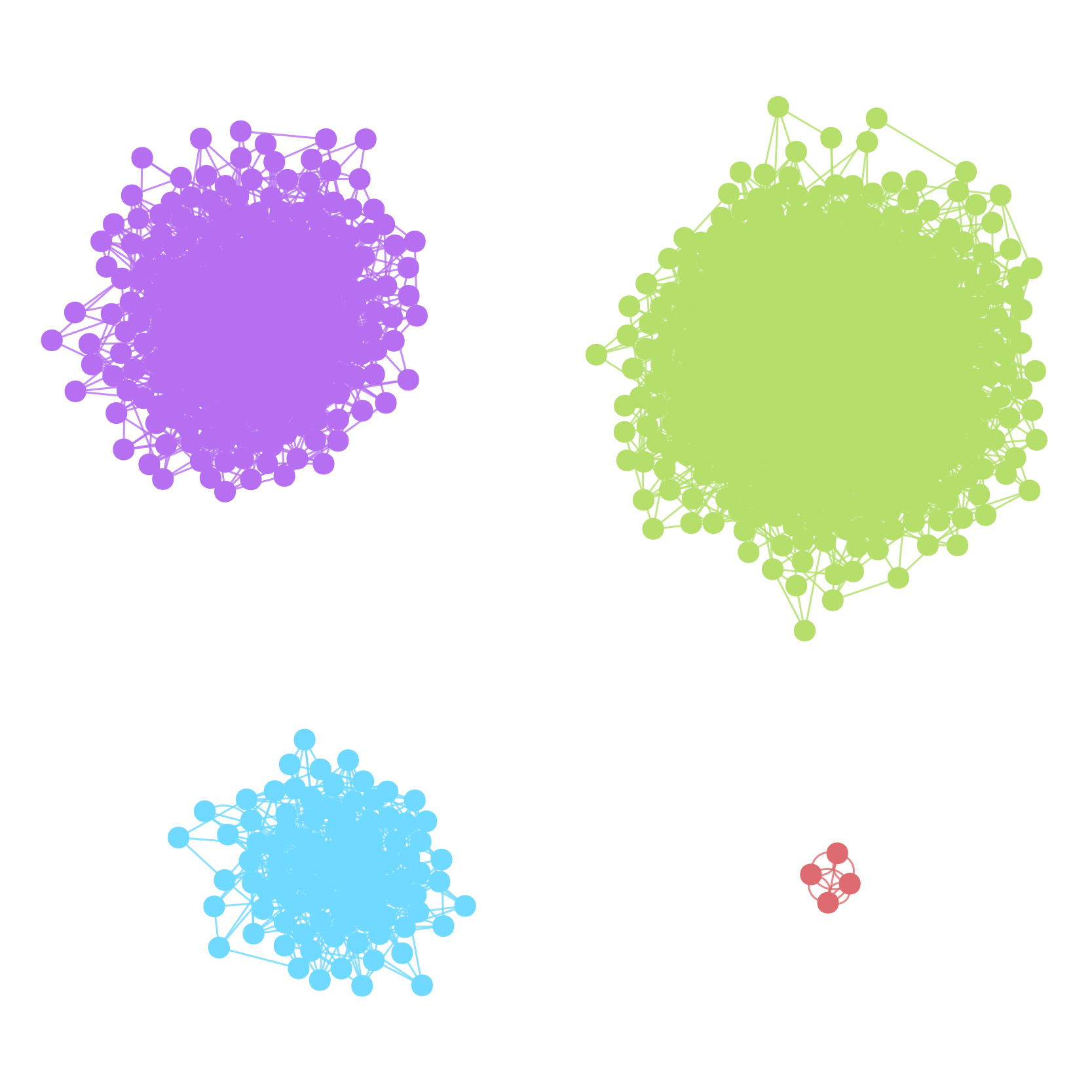}
    \caption{$p_R=0.7$}
\end{subfigure}
\begin{subfigure}[t]{0.3\textwidth}
    \centering
    \includegraphics[width=0.8\textwidth]{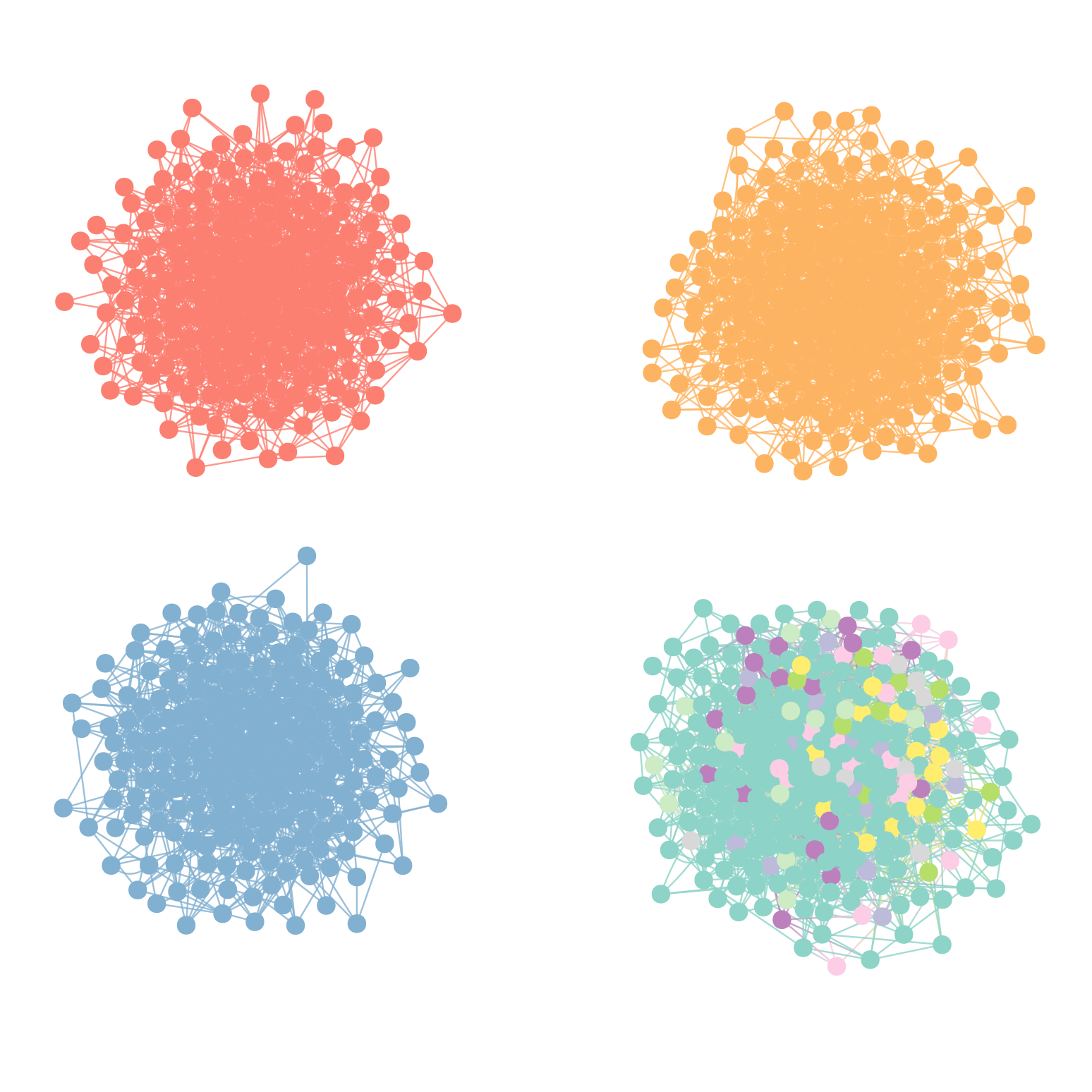}
    \caption{$p_R=0.8$}
\end{subfigure}
\vspace{0.5cm}
\caption{Snapshots of the final network using active rewires, $N=10^3$, $q=10$, $M=4$, $p_N=0$ and varying $p_R$. In (a) we see a simulation that reached consensus before fragmenting, but the rewire dynamics still created a structure the SBM algorithm recognized as a separate community. Increasing $p_R$, we see in (b) a simulation where the different opinions fragmented into components with a very uneven size. Increasing $p_R$ further in (c), the fragmentation happens fast enough that the opinions still hold roughly the same amount of agents}
\label{fig:communities-noiseless}
\end{figure}

The main structure that interests us in the final networks are the presence or absence of communities and how this depends with the different parameters. As previously discussed, the noise will keep the network from fragmenting, so we need a way to tell communities apart inside of a same component. The approach we used was to try to detect communities using only the information of which agents connected with each other (no information about their opinions was used). To this end we used the Stochastic Block Model algorithm of community detection (SBM) implemented in ref \cite{graph-tool}. Some snapshots of the final network in different situations, together with the communities detected by the SBM in them can be found in figure \ref{fig:communities} for the case with noise, while figure \ref{fig:communities-noiseless} shows some noiseless examples.

The SBM can be thought of as a stochastic model for the creation of networks with a predetermined community structure (defined by parameters of the model). From this point of view, SBM community detection works by performing inference to find which parameters would be most likely to output the network we are studying and returning the corresponding community structure. A side effect of this randomness is that the communities that are detected change slightly for different runs of the algorithm, even for the same network. So care must be taken during the analysis to sample multiple runs of the SBM for the same network, besides sampling multiple simulations.

\subsection{Echo chambers}

In order to find out if the dynamics is creating echo chambers we need to cross the information of the communities detected by the SBM with the information of which opinion each agent holds. Suppose we have some community $C$. Define $N(C)$ as the number of agents in this community and $N_{\sigma}(C)$ the number of agents holding opinion $\sigma$ in $C$. We can identify if $C$ has a strong majority opinion by many different metrics, for example:

\begin{equation}
Q_C = \sum_{\sigma} \frac{N_{\sigma}(C)^2}{N(C)^2}
\label{eq:Qc-def}
\end{equation}
or

\begin{equation}
M_C = \max_{\sigma} \left\{\frac{N_{\sigma}(C)}{N(C)}\right\}
\label{eq:Mc-def}
\end{equation}
which can be thought of as proxies for the collision entropy and the min-entropy respectively. It turns out that the results are extremely similar using both metrics, so we only present the results using $Q$. The point is that $Q_C$ is minimum if all opinions appear in $C$ in equal proportions (implying $Q_C = \nicefrac{1}{M}$) and $Q_C$ is maximum if the community contains only one opinion (implying $Q_C = 1$). To extend this to a metric for the whole network we take a weighted average over all communities, using their sizes as weights:

\begin{equation}
Q = \sum_{C} \frac{Q_C N(C)}{N}
\label{eq:Q-def}
\end{equation}

There's two purposes behind using community sizes as weights:
\begin{itemize}
\item We want to minimize the effect that unconnected agents have. The model dynamics leads naturally to a proportion of sites that are not connected to any other, as can be seen in figure \ref{fig:communities}. This situation is temporary as other agents might rewire and connect with them, but there's no good way for the SBM detection to lump these isolated sites with the other communities (since no opinion information is used for the detection) and they are often categorized into small, separated communities.
\item We want to minimize the effect of the SBM detection erroneously splitting a community in two (as seen in some examples in figure \ref{fig:communities}). Since each run of the SBM produces similar, but different communities, it is possible that a community ends up split in two in some SBM runs. If we combine the data using the sizes as weights, the contributions are roughly the same, whether the split happens or not.
\end{itemize}

\pagebreak
\onecolumngrid

\begin{figure}[htbp!]
\centering
\begin{subfigure}[t]{0.48\textwidth}
    \centering
    \includegraphics[width=\textwidth]{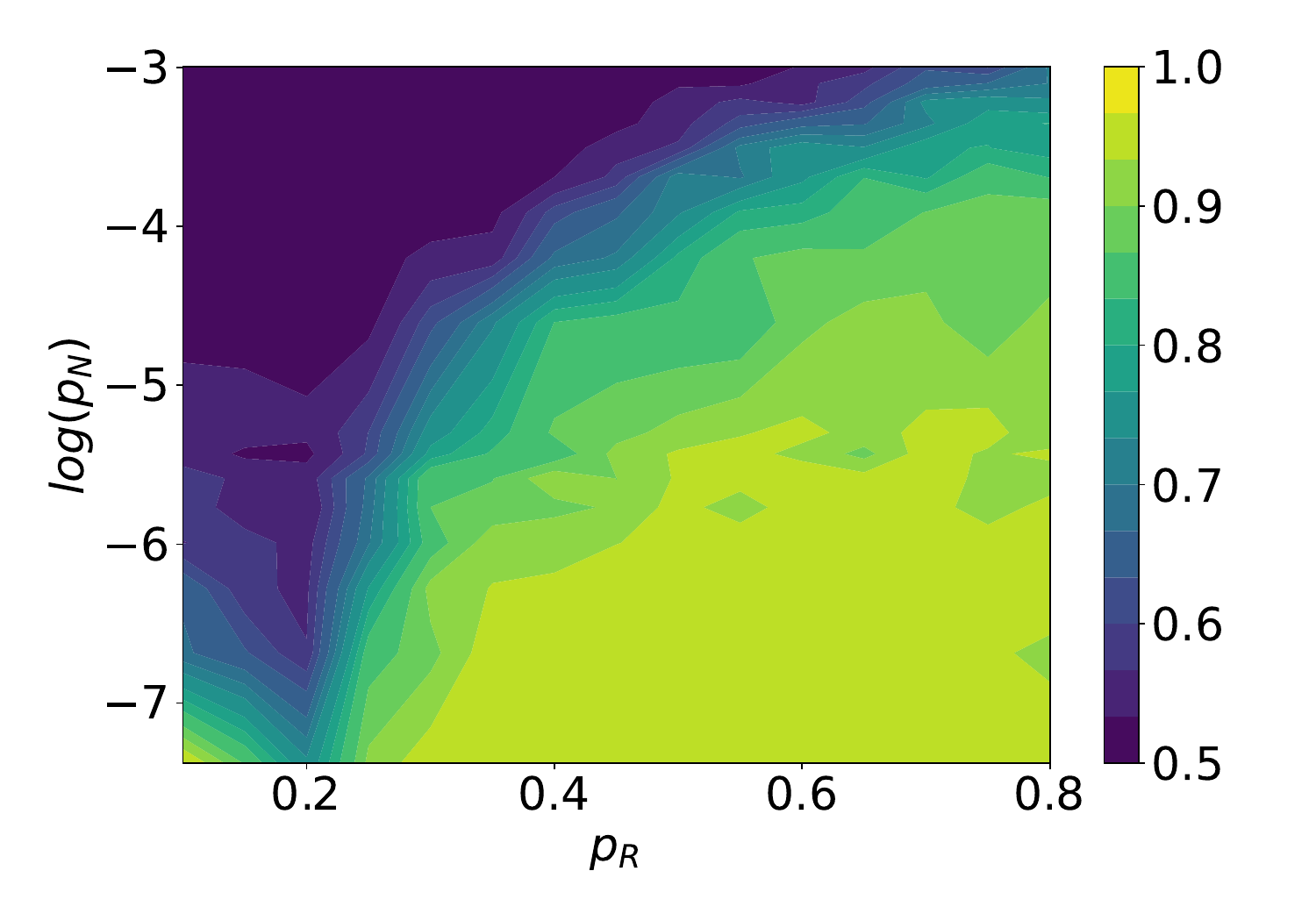}
    \captionsetup{justification=centering}
    \caption{$q = 3$, $M=2$}
\end{subfigure}
\begin{subfigure}[t]{0.48\textwidth}
    \centering
    \includegraphics[width=\textwidth]{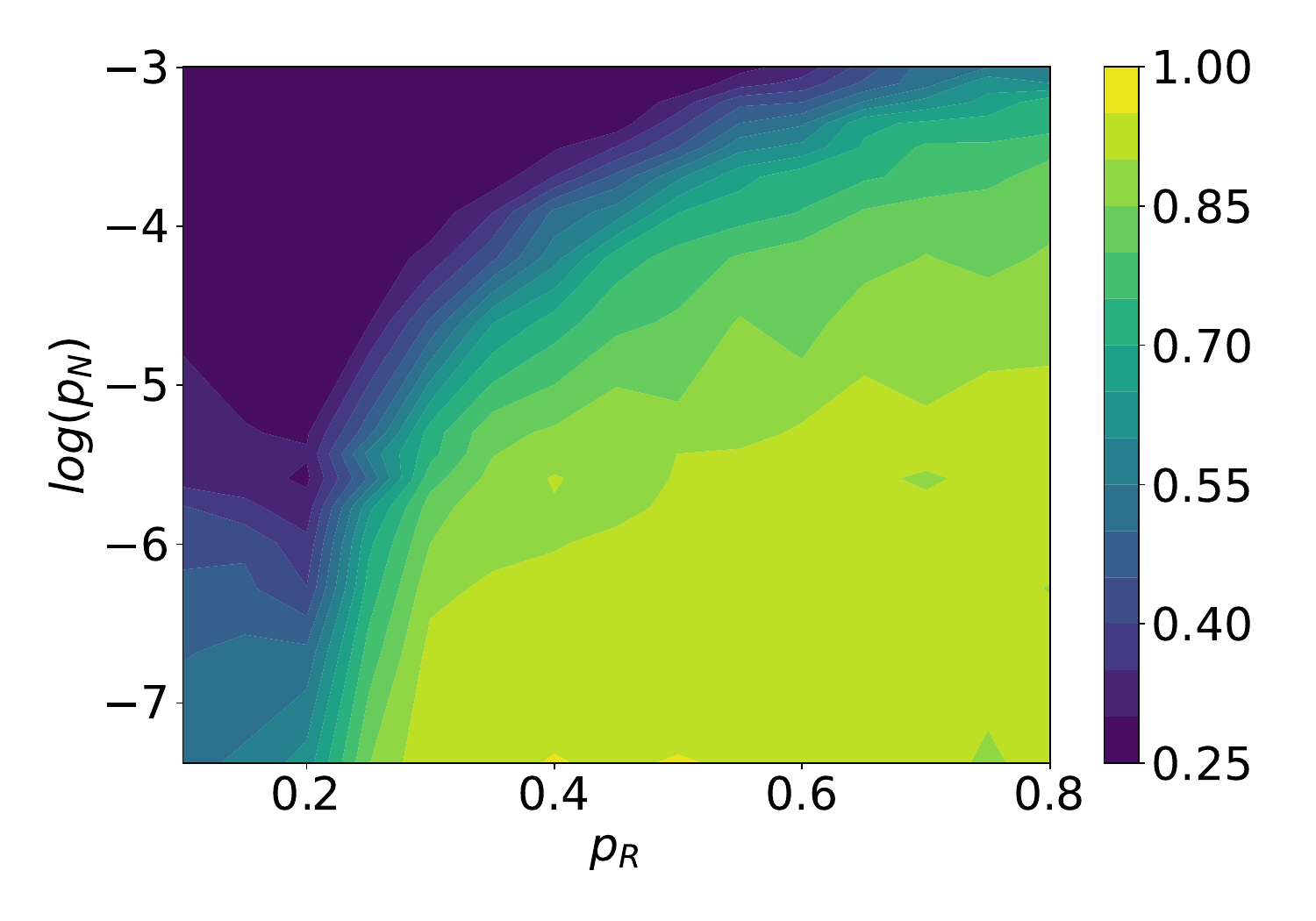}
    \captionsetup{justification=centering}
    \caption{$q = 3$, $M=4$}
\end{subfigure}
\begin{subfigure}[t]{0.48\textwidth}
    \centering
    \includegraphics[width=\textwidth]{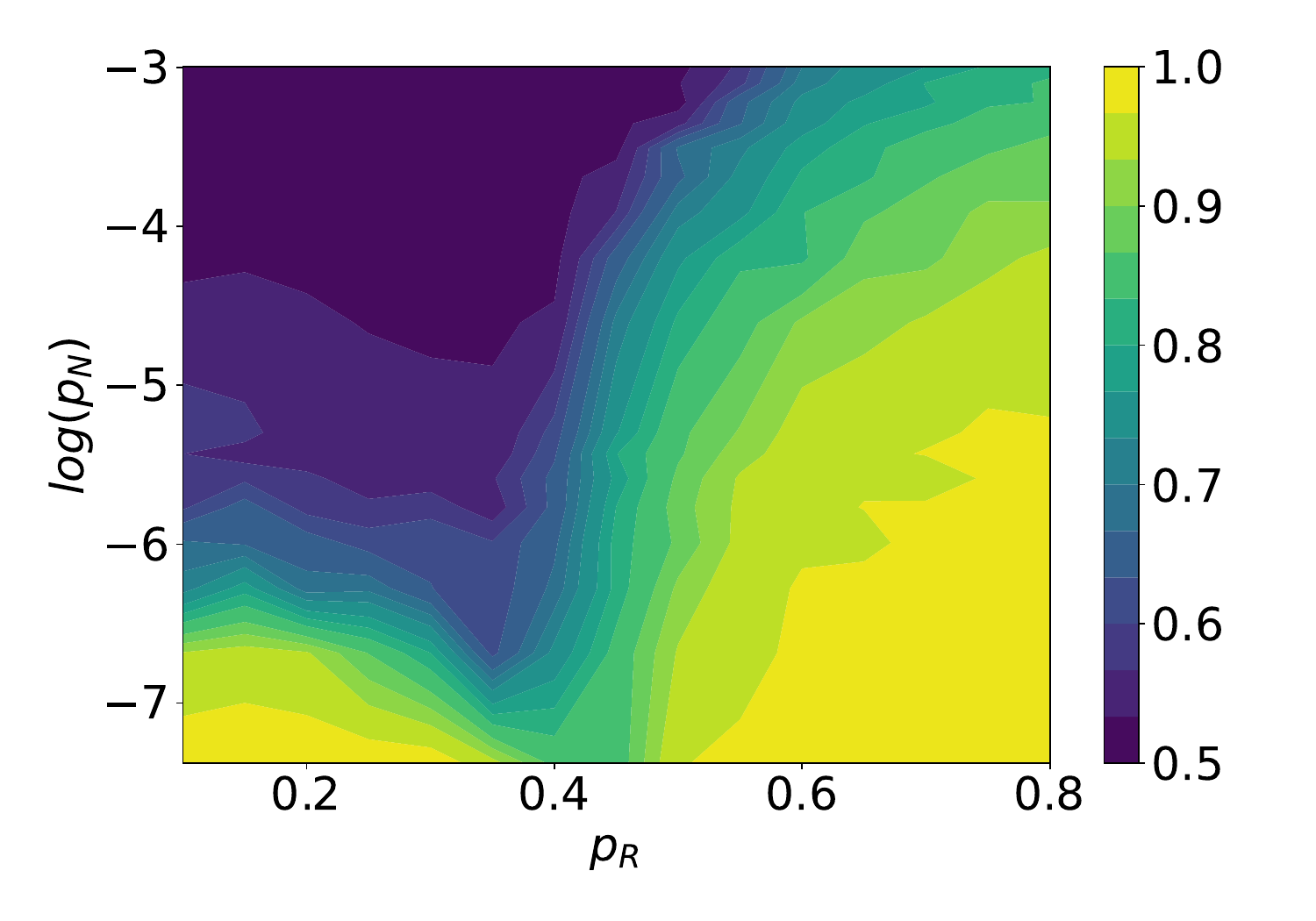}
    \captionsetup{justification=centering}
    \caption{$q = 5$, $M=2$}
    \label{fig:active-example-Q}
\end{subfigure}
\begin{subfigure}[t]{0.48\textwidth}
    \centering
    \includegraphics[width=\textwidth]{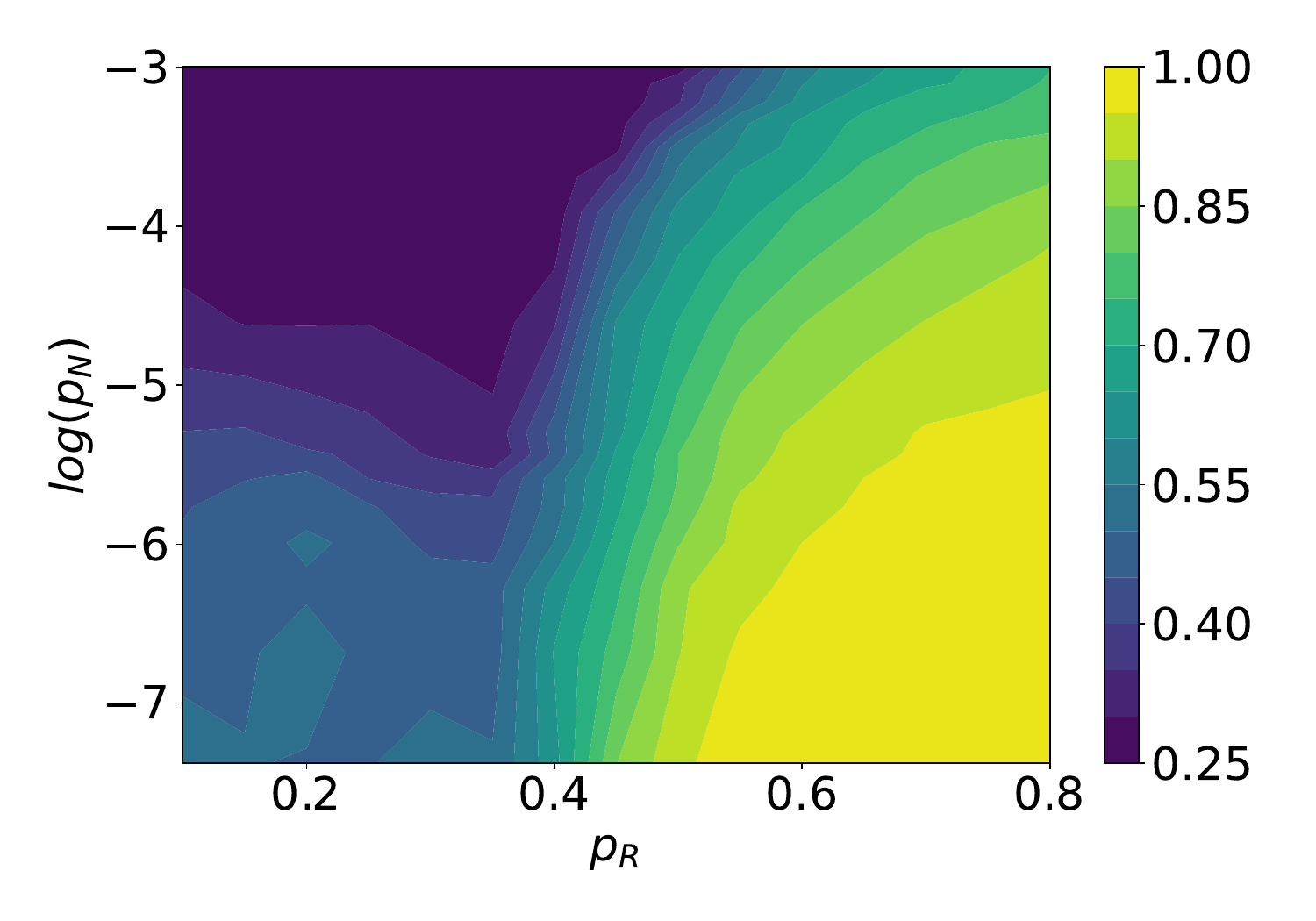}
    \captionsetup{justification=centering}
    \caption{$q = 5$, $M=4$}
\end{subfigure}
\begin{subfigure}[t]{0.48\textwidth}
    \centering
    \includegraphics[width=\textwidth]{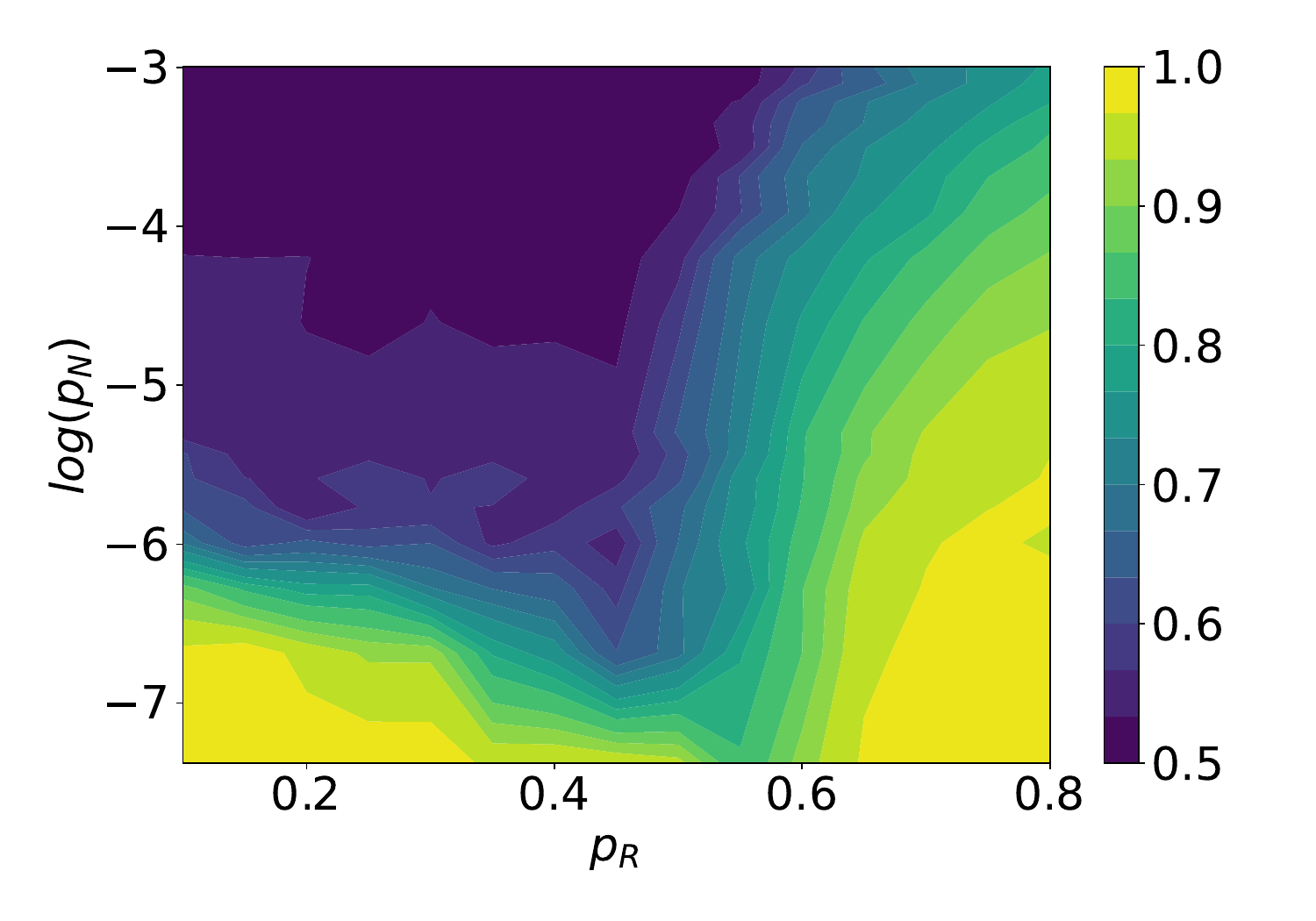}
    \captionsetup{justification=centering}
    \caption{$q = 7$, $M=2$}
\end{subfigure}
\begin{subfigure}[t]{0.48\textwidth}
    \centering
    \includegraphics[width=\textwidth]{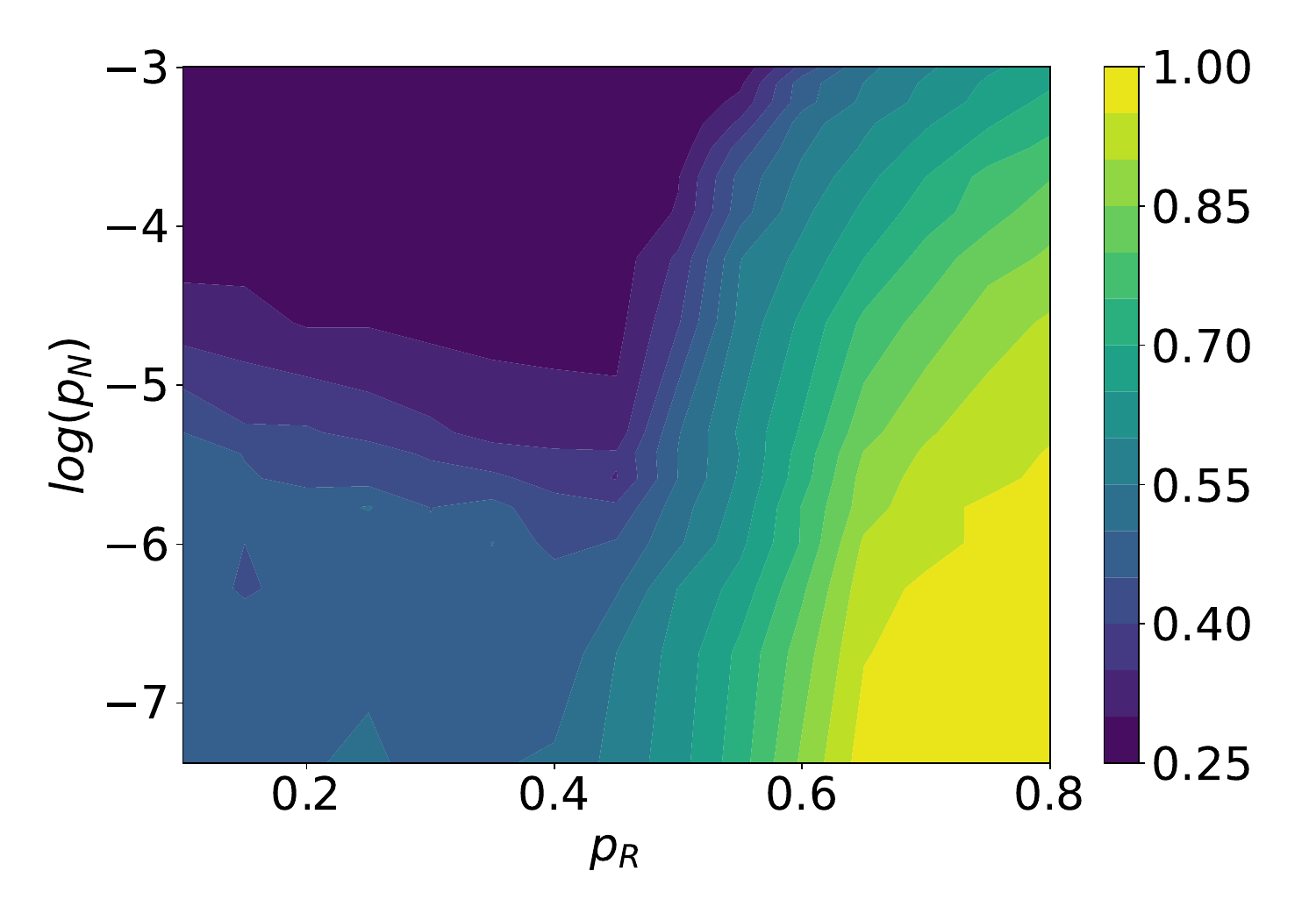}
    \captionsetup{justification=centering}
    \caption{$q = 7$, $M=4$}
\end{subfigure}
\caption{The average $Q$ taken over the communities and averaged over different simulations as a function of the rewiring probability $p_R$ ($x$ axis) and the noise intensity $p_N$ ($y$ axis in natural log for a better view) for different values of $q$ and $M$. Simulations done with active rewiring.}
\label{fig:active-Q-sim}
\end{figure}
\pagebreak

\begin{figure}[htbp!]
\centering
\begin{subfigure}[t]{0.48\textwidth}
    \centering
    \includegraphics[width=\textwidth]{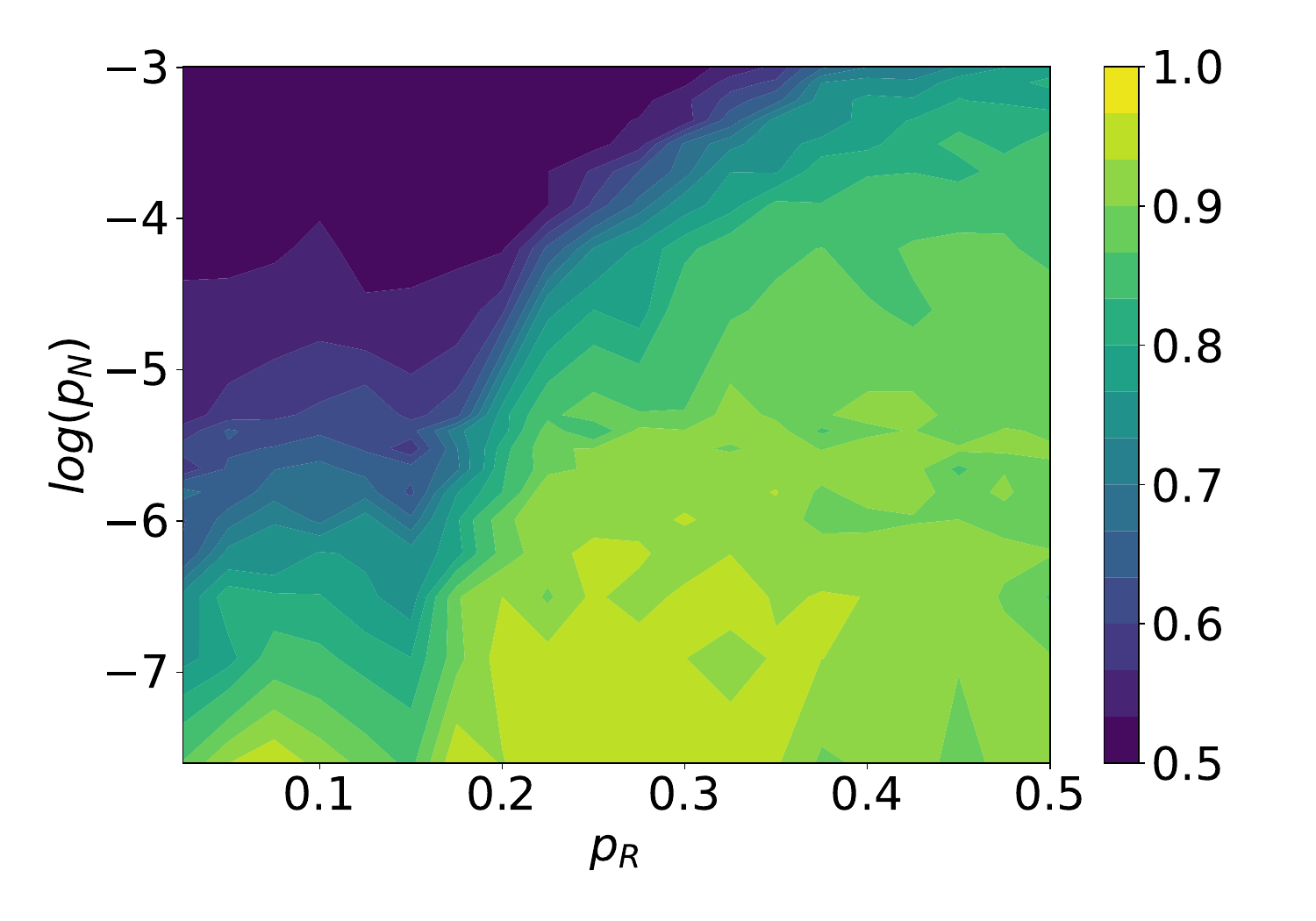}
    \captionsetup{justification=centering}
    \caption{$q = 3$, $M=2$}
\end{subfigure}
\begin{subfigure}[t]{0.48\textwidth}
    \centering
    \includegraphics[width=\textwidth]{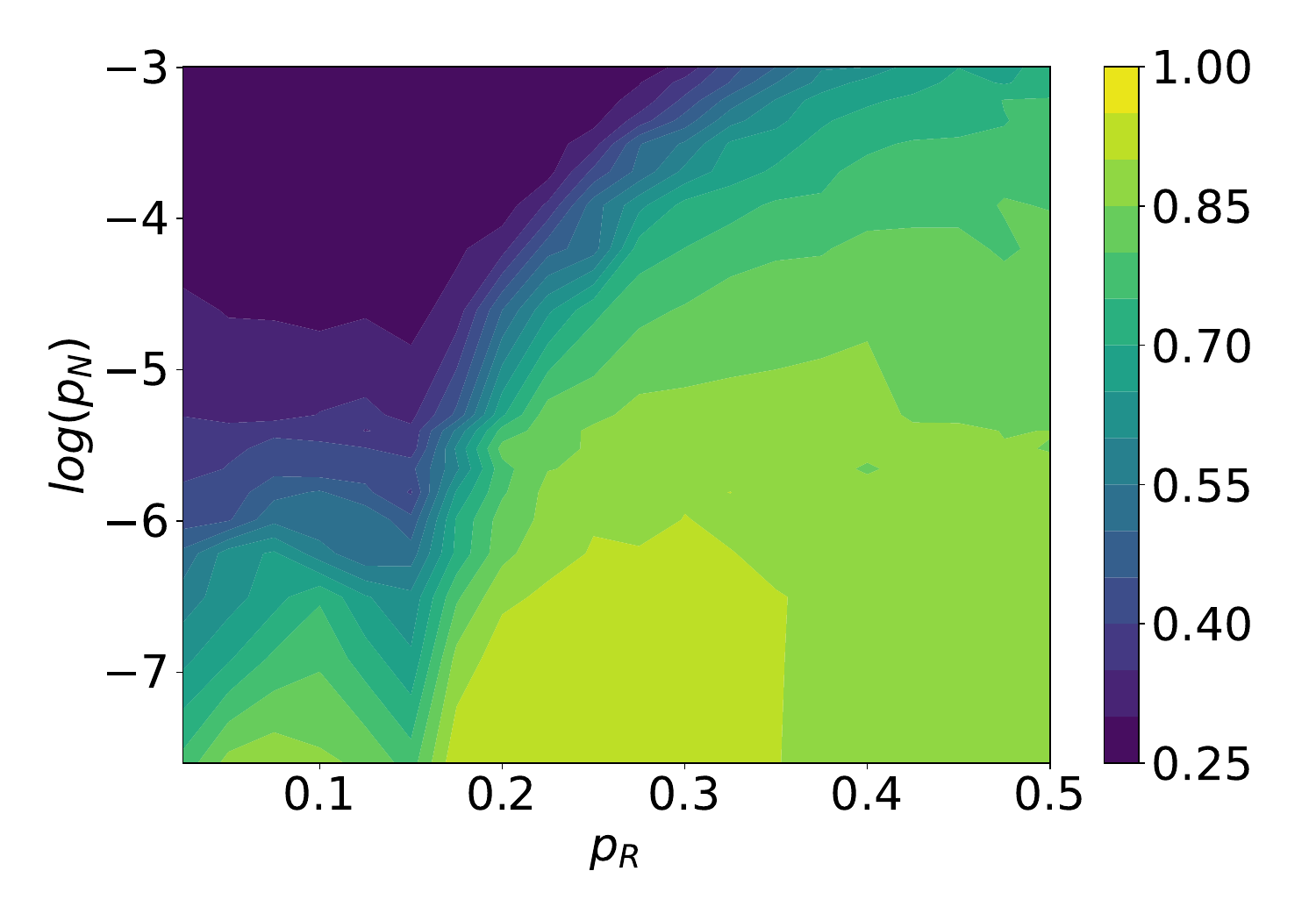}
    \captionsetup{justification=centering}
    \caption{$q = 3$, $M=4$}
\end{subfigure}
\begin{subfigure}[t]{0.48\textwidth}
    \centering
    \includegraphics[width=\textwidth]{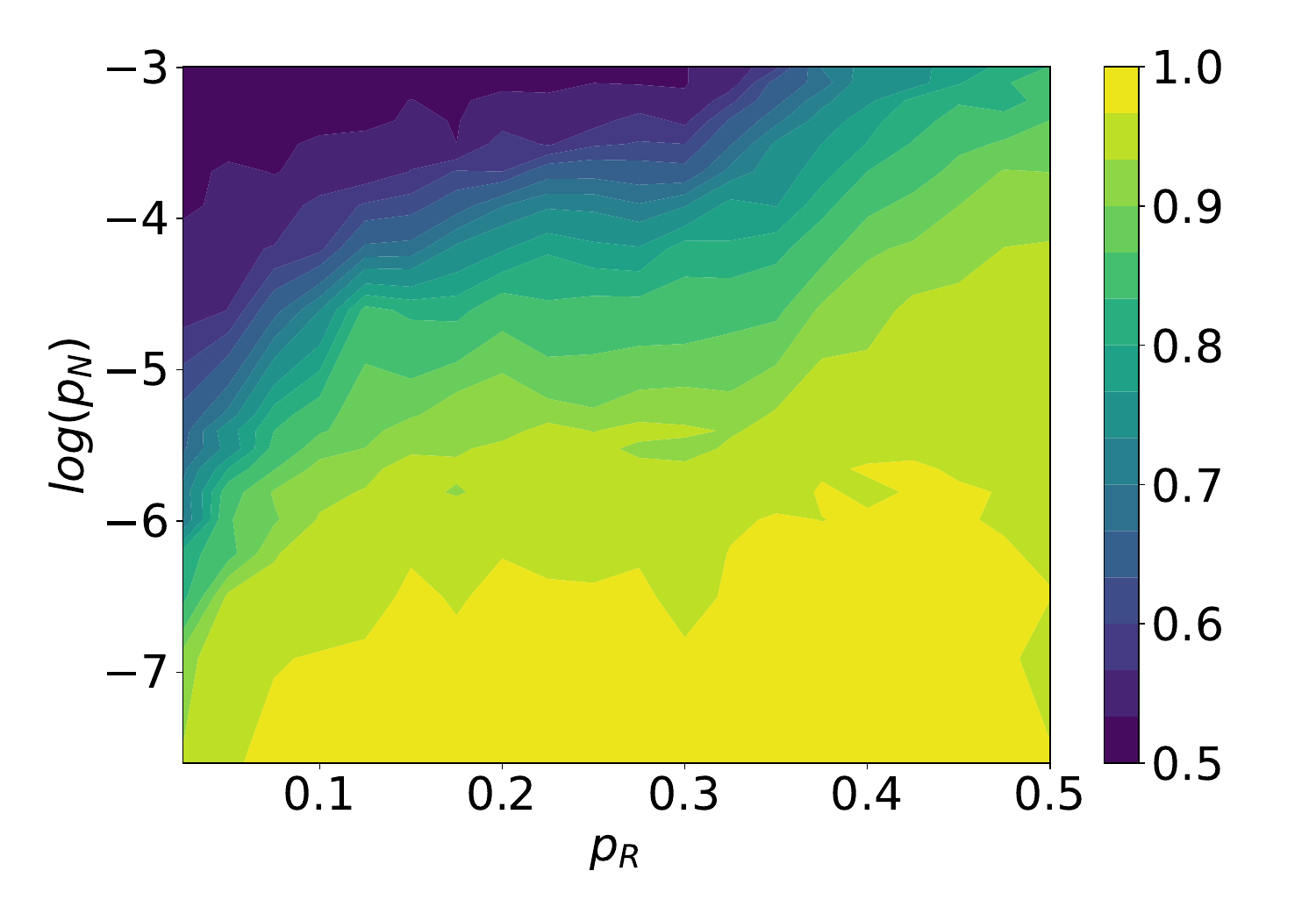}
    \captionsetup{justification=centering}
    \caption{$q = 5$, $M=2$}
\end{subfigure}
\begin{subfigure}[t]{0.48\textwidth}
    \centering
    \includegraphics[width=\textwidth]{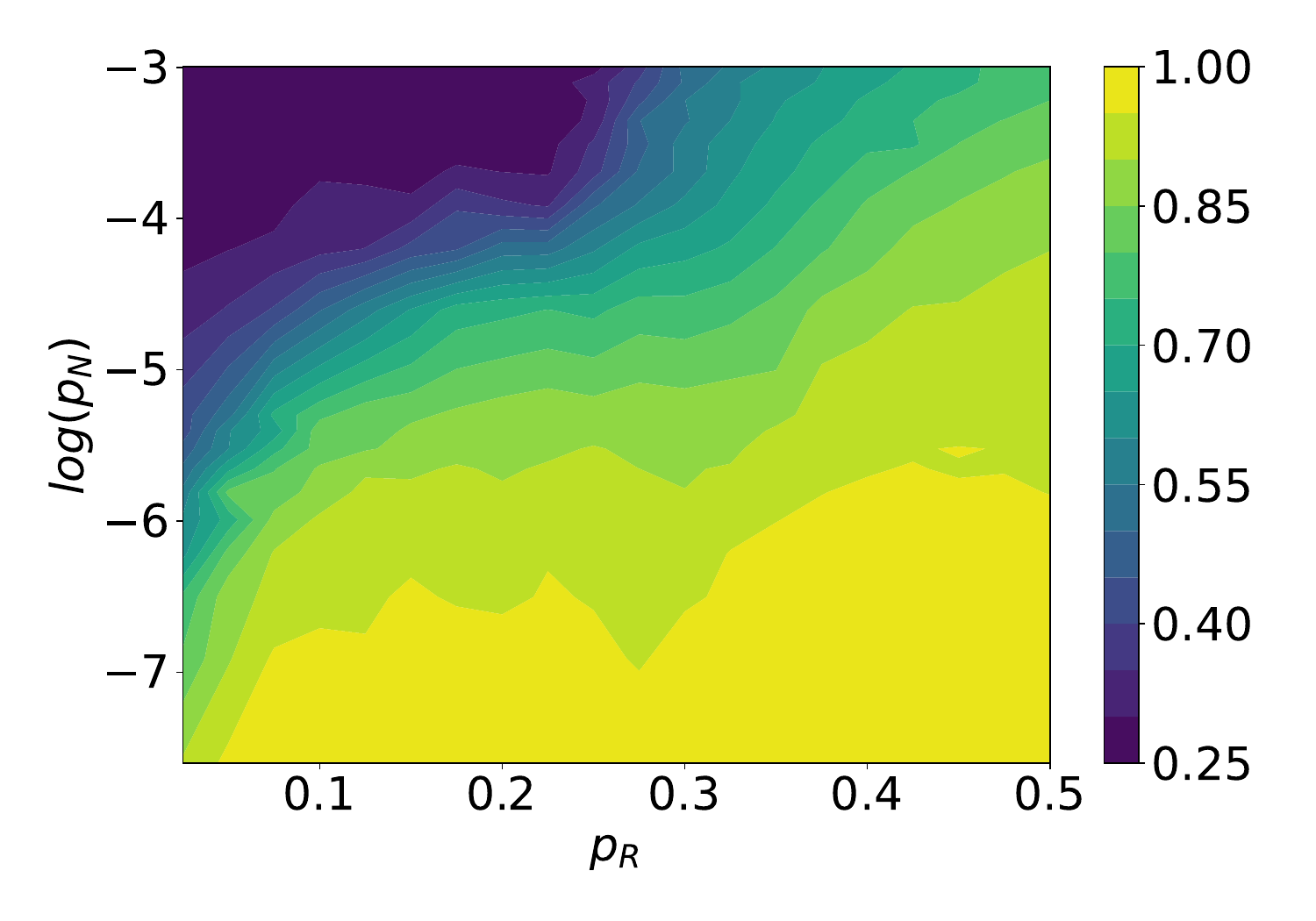}
    \captionsetup{justification=centering}
    \caption{$q = 5$, $M=4$}
\end{subfigure}
\begin{subfigure}[t]{0.48\textwidth}
    \centering
    \includegraphics[width=\textwidth]{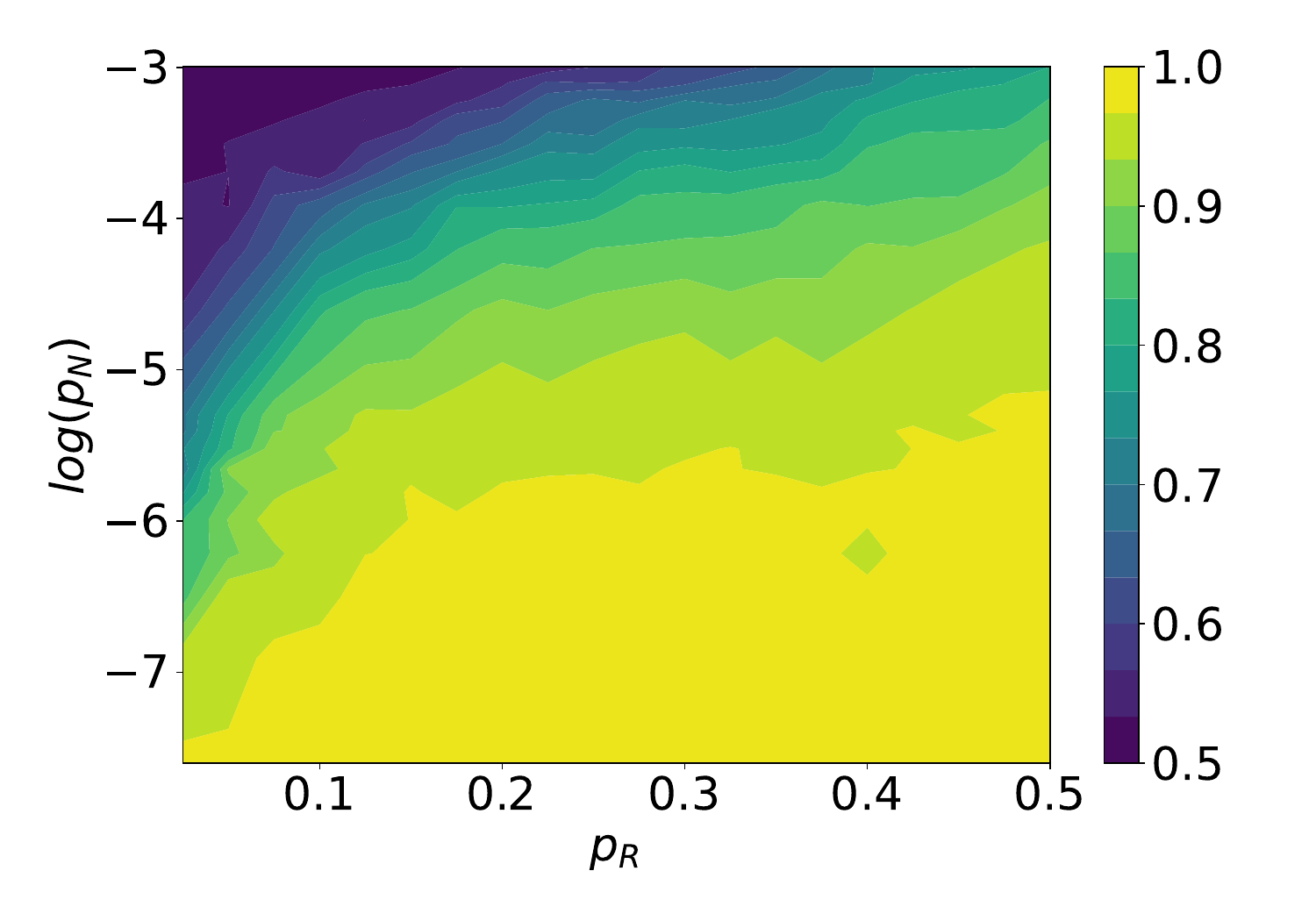}
    \captionsetup{justification=centering}
    \caption{$q = 7$, $M=2$}
\end{subfigure}
\begin{subfigure}[t]{0.48\textwidth}
    \centering
    \includegraphics[width=\textwidth]{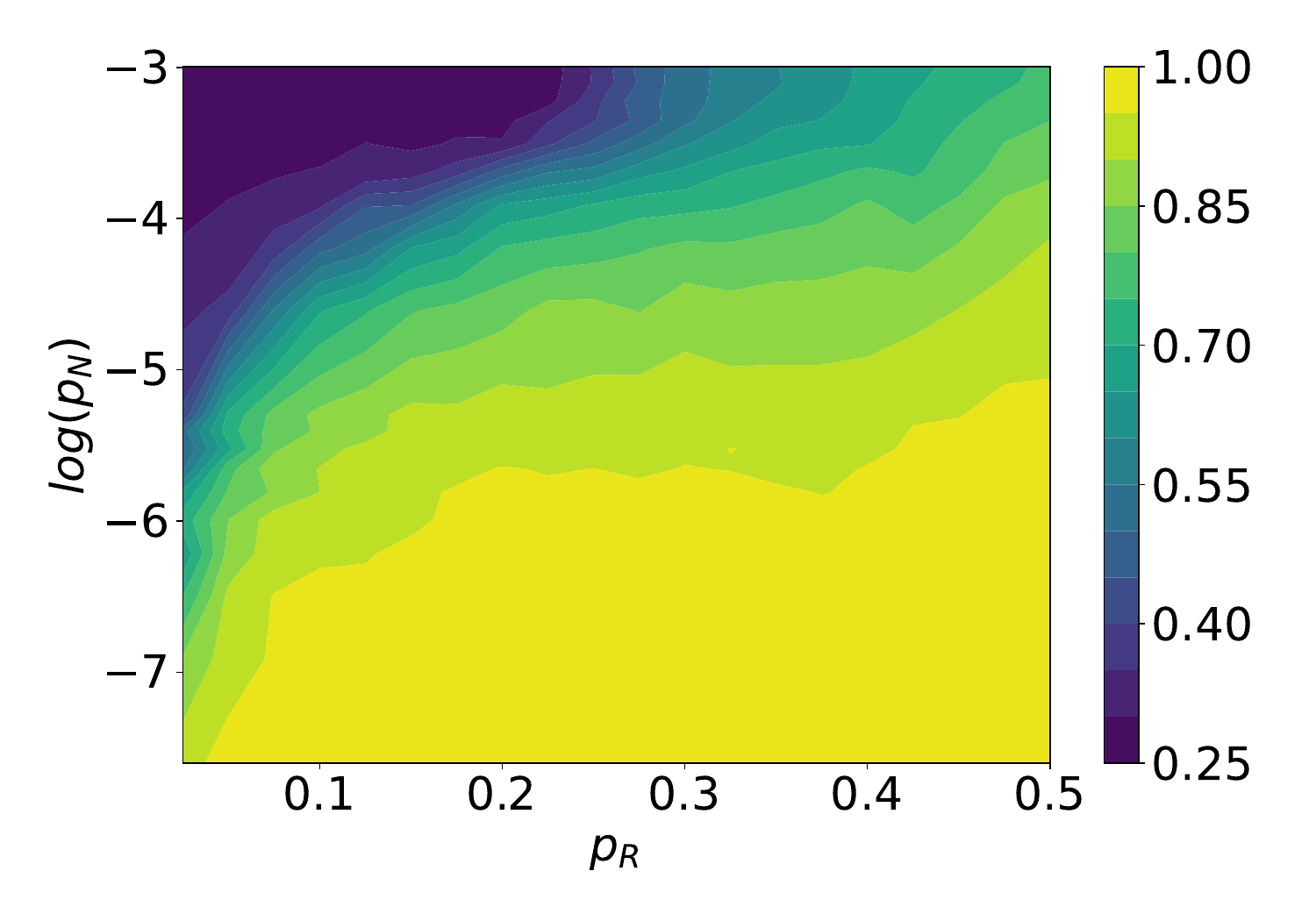}
    \captionsetup{justification=centering}
    \caption{$q = 7$, $M=4$}
\end{subfigure}
\caption{The average $Q$ taken over the communities and averaged over different simulations as a function of the rewiring probability $p_R$ ($x$ axis) and the noise intensity $p_N$ ($y$ axis in natural log for a better view) for different values of $q$ and $M$. Simulations done with reactive rewiring.}
\label{fig:reactive-Q-sim}
\end{figure}
\pagebreak

\twocolumngrid

The graphs of $Q$ for some parameter values, as a function of $p_N$ and $p_R$ can be found in figures \ref{fig:active-Q-sim} (active rewiring) and \ref{fig:reactive-Q-sim} (reactive rewiring). In these graphs we can see a clear transition between regimes where there is a formation of echo chambers (lighter colors) and situations where there is no clear community formation (darker colors). In both cases there's a trend where higher $p_N$ leads to less fragmentation, whereas higher $p_R$ increases fragmentation (corroborating patterns observed in the selected examples of fig \ref{fig:communities}). Interestingly, the number of opinions doesn't seem to play a role in the emergence of not of fragmentation, as the contours in the graphs remain largely the same once we adjust for the different minimum values of $Q$.
In the case of active rewiring a larger $q$ seems to prevent fragmentation better, however this trend seems to be reversed for reactive rewires.

\section{Mean Field Theory}
\label{sec:mf}

\begin{figure}[htbp!]
\centering
\includegraphics[width=0.48\textwidth]{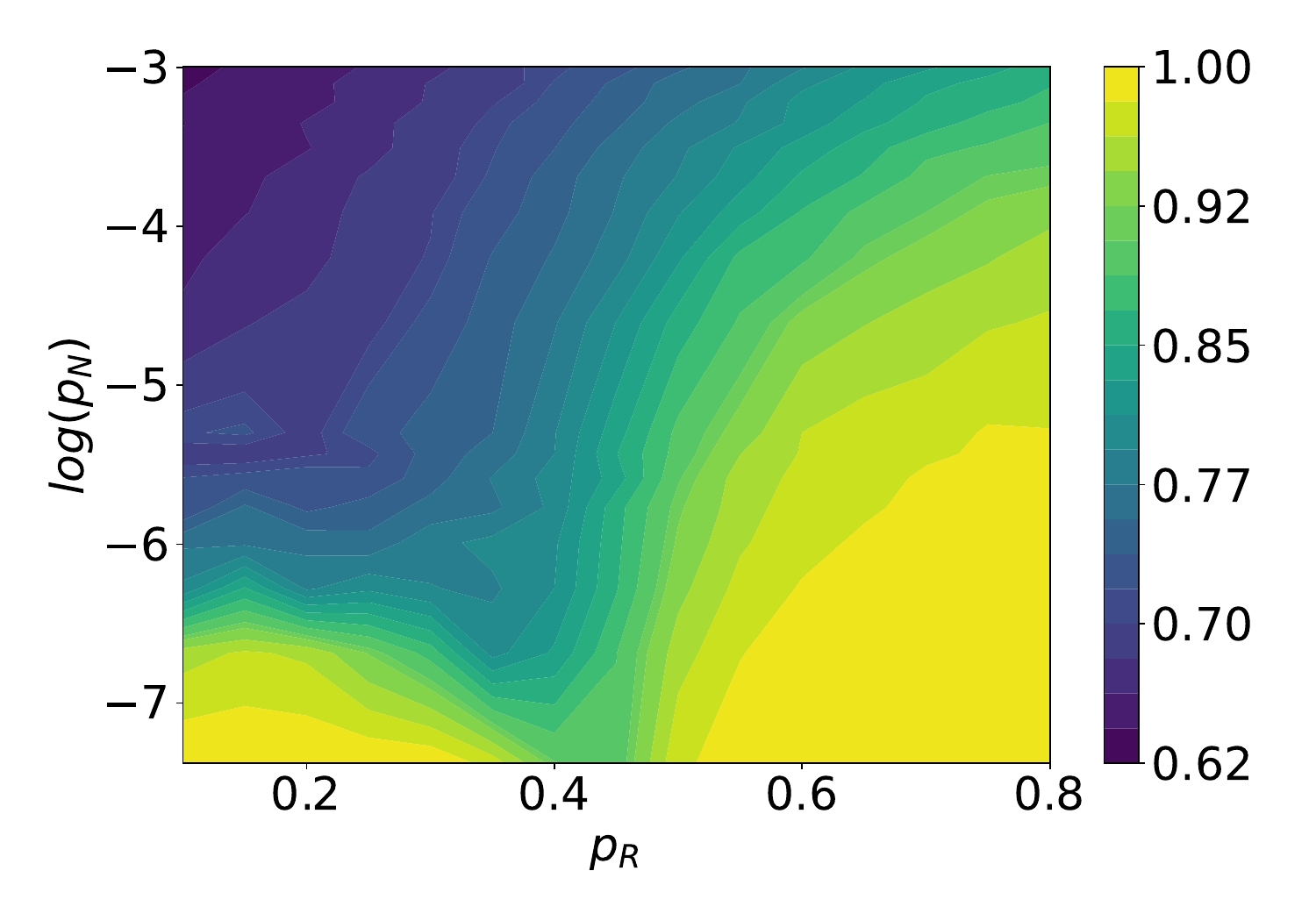}
\captionsetup{justification=centering}
\caption{A graph of $S_{\infty}$ obtained from simulations with active rewiring, $q=5$ and $M=2$ (compare with the graph for $Q$ of the same simulations in fig \ref{fig:active-example-Q})}
\label{fig:S-example}
\end{figure}

Since we are concerned with the community structure, then if we want to do a mean field treatment of this problem we'd need to keep this information about the networks while throwing away the rest of the information. We will achieve this through the following approximations:

\begin{itemize}
\item All opinions are held by the same amount of agents at all times. This is justified on grounds that the noise will keep the system with all opinions being held by about the same amount of agents. So this is akin to neglecting the fluctuations in time.
\item The probability $P_{\sigma, \sigma'}$ that an edge connects agents with opinions $\sigma$ and $\sigma'$ is
\begin{equation}
P_{\sigma, \sigma'} = 
\left\{
\begin{array}{ll}
p_{\mathrm{same}} & \mbox{ if }\sigma = \sigma' \\
p_{\mathrm{different}} & \mbox{ if }\sigma \neq \sigma'
\end{array}
\right.
\label{eq:def-psame}
\end{equation}
With $p_{\mathrm{same}}$ and $p_{\mathrm{different}}$ independent of the specific opinions involved.
\item As seen in figure \ref{fig:deg-dist} the degree distribution converges after a long time. The information from these distributions we will need for our calculations are the average $q$ and the probability $P_0$ of finding an isolated agent. We will assume $P_0$ is a known parameter of our mean field theory.
\end{itemize}
With these approximations in place, then the community structure is entirely given by the parameters $N$, $M$, $q$, $P_0$ and the probability $p_{\mathrm{same}}$ ($p_{\mathrm{different}}$ can be obtained as a function of $p_{\mathrm{same}}$ with the equation $Mp_{\mathrm{same}} + M(M-1)p_{\mathrm{different}} = 1$). It will be more convenient to use the probability that a pair of neighbouring agents have the same opinion, which we will denote simply by $S$ ($S = M p_{\mathrm{same}}$). This is a quantity that can also be measured in the regular simulations and also gives us a proxy to the formation of echo chambers, as exemplified by figure \ref{fig:S-example}

Our goal with the mean field treatment is to understand how $S$ evolves in time (under the assumption that our approximations are valid at all times) to obtain
\begin{equation}
    S_{\infty} = \lim_{t\rightarrow \infty} S(t)
    \label{eq:p-lim}
\end{equation}
and compare the structure that this implies with our simulation results.

\subsection{Mean field evolution}
The idea of the mean field evolution is to make a time step using our approximations and seeing how this timestep changes $S$. Following the model descriptions in sections \ref{sec:active-rewires} and \ref{sec:reactive-rewires}, the mean field timesteps look like:

\begin{itemize}
\item Check if the agent $i$ to be chosen will be affected by the noise.
\item Draw from the appropriate distributions the number of neighbours of $i$ and how many of these neighbours agree with $i$.
\item If $i$ is not isolated, choose a neighbour $j$ for $i$ to interact with.
\item Do the appropriate rewires along the evolution.
\end{itemize}

However, instead of making a simulation, we analytically find how $S$ would change after this timestep by adding together all possible outcomes with the appropriate weights. The detailed calculations of these averages can be found in appendix \ref{app:details}. In the interest of making the final expressions more compact we define the quantities: $p_N^{\complement} = 1-p_N$, $p_R^{\complement} = 1-p_R$, $P_0^{\complement} = 1-P_0$ and $S^{\complement} = 1-S$

\subsubsection{Active rewires}
The detailed timestep for active rewires is as follows:

\begin{itemize}
    \item With probability $p_N$ (agent $i$ is affected by the noise):
    \begin{itemize}
        \item With probability $\nicefrac{1}{M}$ (the noise doesn't change the opinion of $i$):
        \begin{itemize}
            \item Nothing happens\vspace{0.1cm}
        \end{itemize}
        \item Otherwise (probability $\nicefrac{(M-1)}{M}$):
        \begin{itemize}
            \item We draw the number of neighbours $k$ from the degree distribution\vspace{0.2cm}
            \item We draw $I\sim $ \,Binomial$(N=k,p=S)$, the number of neighbours agreeing with the agent's previous opinion. \vspace{0.2cm}
            \item We draw $n\sim $ \,Binomial$(N=k-I,p=\nicefrac{1}{(M-1)})$, the number of neighbours that agree with the agent's new opinion. \vspace{0.2cm}
            \item $S$ changes by $\nicefrac{(n - I)}{E}$, where $E = \nicefrac{Nq}{2}$ is the number of edges in the network. On average, this contribution is
            \begin{equation}
                \Delta S^{(a)}_1 = \frac{p_N q (1-M S)}{EM}
                \label{eq:contrib1}
            \end{equation}
        \end{itemize}
    \end{itemize}
    \item Otherwise (probability $1-p_N$), if agent $i$ is not affected by the noise:
    \begin{itemize}
        \item We draw the number of neighbours $k$ from the degree distribution \vspace{0.2cm}
        \item We draw $I\sim $ \,Binomial$(N=k,p=S)$, the number of neighbours agreeing with the agent's current opinion. \vspace{0.2cm}
        \item If $k = 0$ or otherwise with probability $\nicefrac{I}{k}$ (agent $i$ is either isolated or interacts with a neighbour that it agrees with):
        \begin{itemize}
            \item Nothing happens \vspace{0.1cm}
        \end{itemize}
        \item Otherwise (agent $i$ interacts with a neighbour $j$ it doesn't agree with):
        \begin{itemize}
            \item With probability $p_R$ (a rewire happens):\vspace{0.1cm}
            \begin{itemize}
                \item This changes $S$ by $\nicefrac{1}{E}$. On average this contribution is
                \begin{equation}
                    \Delta S^{(a)}_2 = \frac{p_R \,p_N^{\complement}S^{\complement}P_0^{\complement}}{E}
                    \label{eq:contrib2}
                \end{equation}
            \end{itemize}
            \item Otherwise (probability $1 - p_R$) the agent copies its neighbour:\vspace{0.1cm}
            \begin{itemize}
                \item We draw $n\sim $ \,Binomial$(N=k-I-1,p=\nicefrac{1}{(M-1)})$ the number of neighbours of agent $i$, besides $j$ that agree with the agent's new opinion.\vspace{0.2cm}
                \item $S$ changes by $\nicefrac{(n + 1 - I)}{E}$. On average this contribution is
                \begin{equation}
                    \Delta S^{(a)}_3 = \frac{p_R^{\complement}p_N^{\complement}S^{\complement}((MS + M -2)P_0^{\complement} + q(1 - MS))}{E(M-1)}
                    \label{eq:contrib3}
                \end{equation}
            \end{itemize}
        \end{itemize}
    \end{itemize}
\end{itemize}
So the mean field time evolution is given by (measuring time in Monte Carlo timesteps)
\begin{equation}
    S\left(t + \frac{1}{N}\right) = S(t) + \Delta S^{(a)}_1 + \Delta S^{(a)}_2 + \Delta S^{(a)}_3
    \label{eq:MF-active}
\end{equation}
that becomes an ODE in the limit $N\rightarrow \infty$.

\subsubsection{Reactive rewires}
The detailed timestep for reactive rewires is as follows:

\begin{itemize}
    \item With probability $p_N$ (agent $i$ is affected by the noise):
    \begin{itemize}
        \item With probability $\nicefrac{1}{M}$ (the noise doesn't change the opinion of $i$):
        \begin{itemize}
            \item Nothing happens\vspace{0.1cm}
        \end{itemize}
        \item Otherwise (probability $\nicefrac{(M-1)}{M}$):
        \begin{itemize}
            \item We draw the number of neighbours $k$ from the degree distribution\vspace{0.2cm}
            \item We draw $I\sim $ \,Binomial$(N=k,p=S)$, the number of neighbours agreeing with the agent's previous opinion. \vspace{0.2cm}
            \item We draw $n\sim $ \,Binomial$(N=k-I,p=\nicefrac{1}{(M-1)})$, the number of neighbours that agree with the agent's new opinion. \vspace{0.2cm}
            \item We draw $r\sim $ \,Binomial$(N=k-n,p=p_R)$, the number of neighbours that rewire their connection with $i$. \vspace{0.2cm}
            \item $S$ changes by $\nicefrac{(n + r - I)}{E}$. On average, this contribution is
            \begin{equation}
                \Delta S^{(r)}_1 = \frac{p_N q(M p_R - MS + Sp_R - 2p_R + 1)}{EM}
                \label{eq:contrib4}
            \end{equation}
        \end{itemize}
    \end{itemize}
    \item Otherwise (probability $1-p_N$), if agent $i$ is not affected by the noise:
    \begin{itemize}
        \item We draw the number of neighbours $k$ from the degree distribution \vspace{0.2cm}
        \item We draw $I\sim $ \,Binomial$(N=k,p=S)$, the number of neighbours agreeing with the agent's current opinion. \vspace{0.2cm}
        \item If $k = 0$ or otherwise with probability $\nicefrac{I}{k}$ (agent $i$ is either isolated or interacts with a neighbour that it agrees with):
        \begin{itemize}
            \item Nothing happens \vspace{0.1cm}
        \end{itemize}
        \item Otherwise (agent $i$ interacts with a neighbour $j$ it doesn't agree with):
        \begin{itemize}
            \item $i$ copies $j$'s opinion.\vspace{0.2cm}
            \item We draw $n\sim $ \,Binomial$(N=k-I-1,p=\nicefrac{1}{(M-1)})$ the number of neighbours of agent $i$, besides $j$ that agree with the agent's new opinion.\vspace{0.2cm}
            \item We draw $r\sim $ \,Binomial$(N=k-n-1,p=p_R)$, the number of neighbours that rewire their connection with $i$. \vspace{0.2cm}
            \item $S$ changes by $\nicefrac{(n + r + 1 - I)}{E}$. On average this contribution is
            \begin{IEEEeqnarray}{rCl}\label{eq:contrib5}
                \Delta S^{(r)}_2 = \left(\phantom{\frac{1}{1}}\!\!\!\!\!\!\right.(M(p_R^{\complement} & + & S) - 2p_R^{\complement} - S p_R)(P_0^{\complement} - q) + 
                \notag\\[0.2cm]
                & + & \left. q(M -1)\middle)\frac{p_N^{\complement}S^{\complement}}{E(M - 1)} \right.
            \end{IEEEeqnarray}
        \end{itemize}
    \end{itemize}
\end{itemize}
So the mean field time evolution is given by (measuring time in Monte Carlo timesteps)
\begin{equation}
    S\left(t + \frac{1}{N}\right) = S(t) + \Delta S^{(r)}_1 + \Delta S^{(r)}_2
    \label{eq:MF-reactive}
\end{equation}
that also becomes an ODE in the limit $N\rightarrow \infty$.

\subsection{Long time behaviour}
\subsubsection{Qualitative analysis}
Both equations (\ref{eq:MF-active}) and (\ref{eq:MF-reactive}) reduce when $N\rightarrow \infty$ to an ODE with form

\begin{equation}
\frac{dS}{dt} = \frac{2\theta_{a(r)}(S)}{q}
\label{eq:EDO-quali}
\end{equation}
where $\theta_a$ and $\theta_r$ (for active and reactive rewires respectively) are quadratic. As such (\ref{eq:EDO-quali}) can have at most 2 fixed points.

Examining $\theta$ for $S = 0, 1$ and assuming $M \geq 2$, $q > 0$, $p_N > 0$ and $p_R > 0$, we get the following inequalities:

\begin{equation}
    \theta_a(0) = p_RP_0^{\complement}p_N^{\complement} + \frac{p_N^{\complement}p_R^{\complement}((M - 2)P_0^{\complement} + q)}{M - 1} + \frac{p_Nq}{M} > 0
    \label{eq:qa0}
\end{equation}
\begin{equation}
    \theta_a(1) = \frac{-p_N \,q (M - 1)}{M} < 0
    \label{eq:qa1}
\end{equation}

\begin{IEEEeqnarray}{rCl}\label{eq:qr0}
    \theta_r(0) & = & \frac{p_N^{\complement}(P_0^{\complement}(M-2)p_R^{\complement} + q((M-2)p_R + 1))}{M-1} + 
    \notag\\[0.2cm]
    & & + \frac{p_N \,q (p_R(M-2) + 1)}{M} > 0
\end{IEEEeqnarray}
\begin{equation}
    \theta_r(1) = \frac{-p_N \,q \,p_R^{\complement} (M - 1)}{M} < 0
    \label{eq:qr1}
\end{equation}
where the inequality in equation (\ref{eq:qr1}) further assumes $p_R \neq 1$. Equations (\ref{eq:qa0}) to (\ref{eq:qr1}) imply that there is exactly one fixed point of (\ref{eq:EDO-quali}) with $0 \leq S \leq 1$ and it must be attractive, so the long time behaviour of $S$, $S_{\infty}$ (defined in equation (\ref{eq:p-lim})) can be obtained simply solving $\theta(S) = 0$ with $S$ being a valid probability.

\subsubsection{Comparison with simulations}

In order to compare the mean field calculations with $S$ measured from the simulations, note that the mean field results will still depend on the network geometry, since $P_0$ is still a parameter. In the case of active rewiring, the similarities between the stationary degree distribution and the degree distribution of an Erdös-Rényi network (as evidenced by figure \ref{fig:deg-dist}) suggest that a Poisson distribution might be a good approximation, which would lead to $P_0 \sim e^{-q}$. However, since no such approximation is clear in the reactive case, we opted for using a value of $P_0$ obtained from simulations in both cases.

\begin{figure}[htb!]
\centering
\begin{subfigure}[t]{0.4\textwidth}
    \centering
    \includegraphics[width=\textwidth]{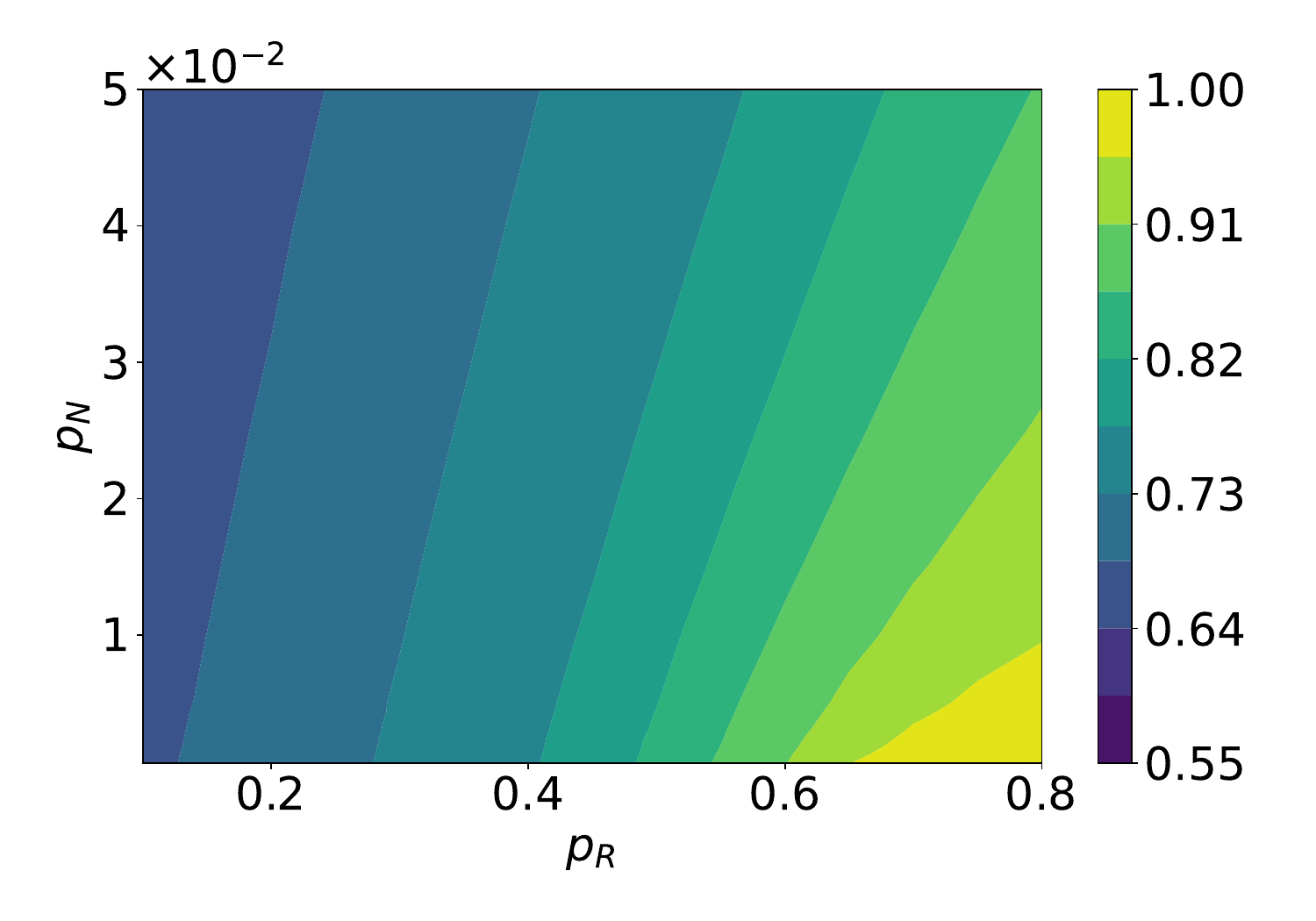}
    \captionsetup{justification=centering}
    \caption{Mean Field with active rewiring, $q=4$ and $M=2$}
\end{subfigure}
\begin{subfigure}[t]{0.4\textwidth}
    \centering
    \includegraphics[width=\textwidth]{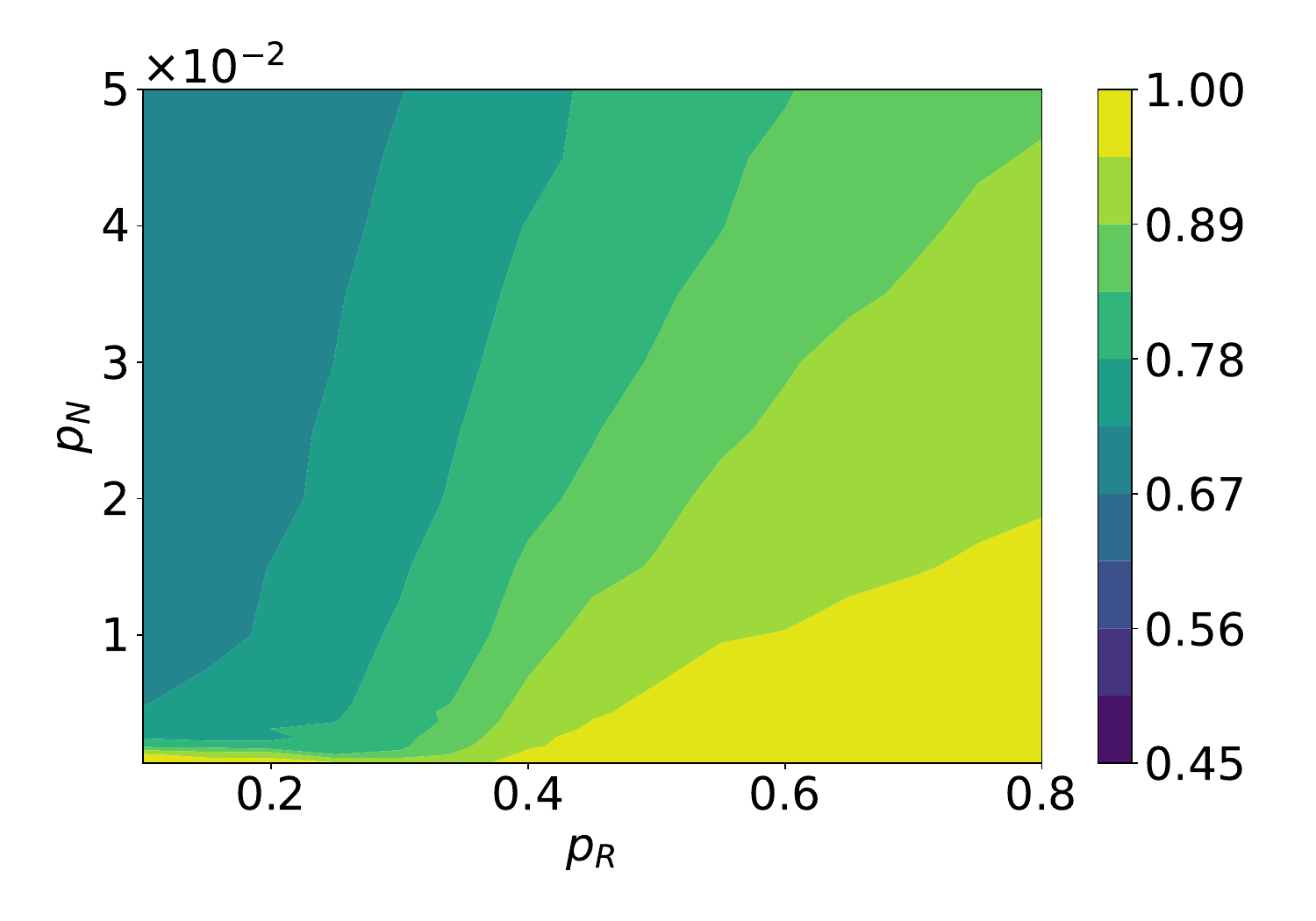}
    \captionsetup{justification=centering}
    \caption{Simulations with active rewiring, $q=4$ and $M=2$}
\end{subfigure}
\begin{subfigure}[t]{0.4\textwidth}
    \centering
    \includegraphics[width=\textwidth]{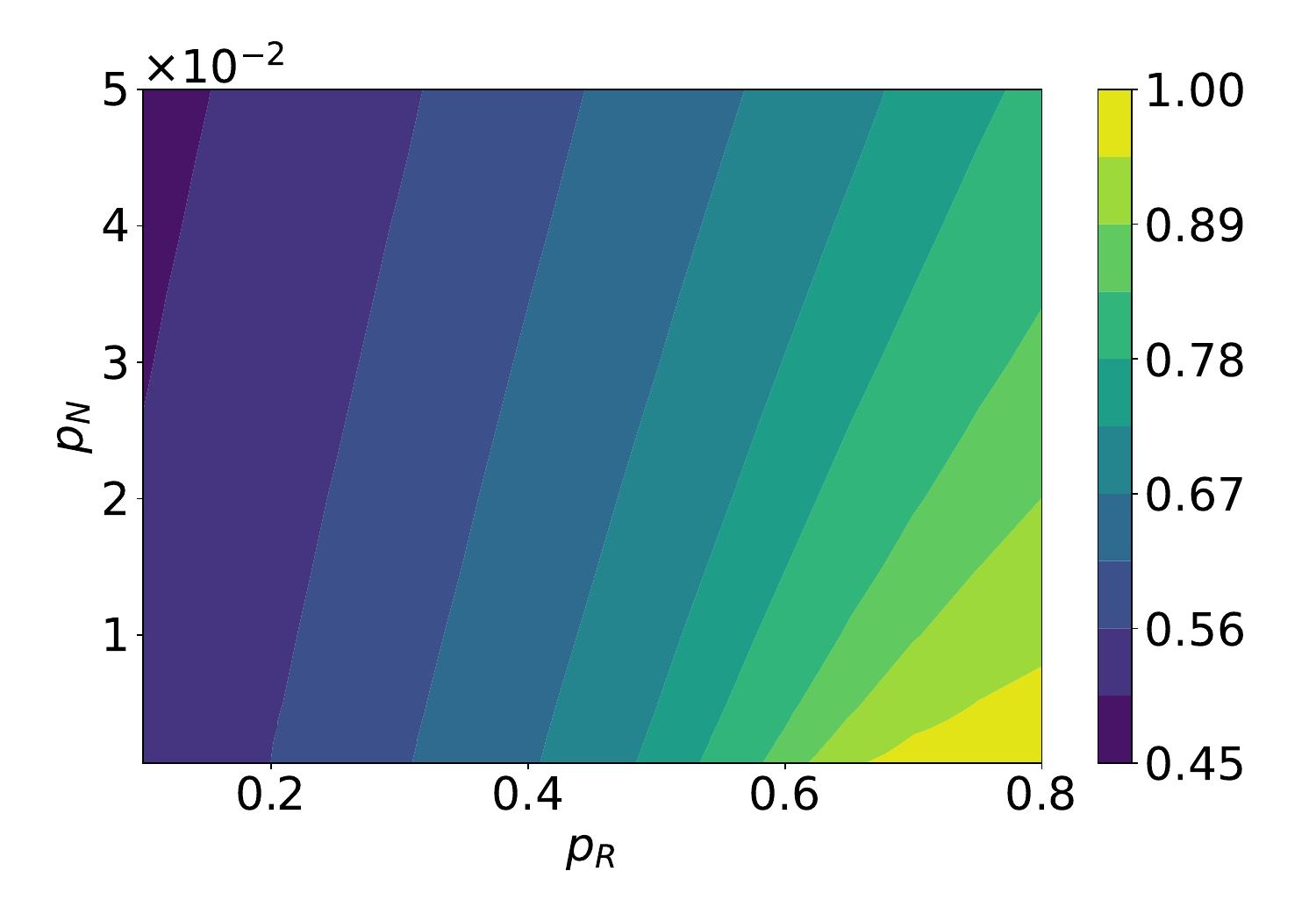}
    \captionsetup{justification=centering}
    \caption{Mean Field with active rewiring, $q=4$ and $M=4$}
\end{subfigure}
\begin{subfigure}[t]{0.4\textwidth}
    \centering
    \includegraphics[width=\textwidth]{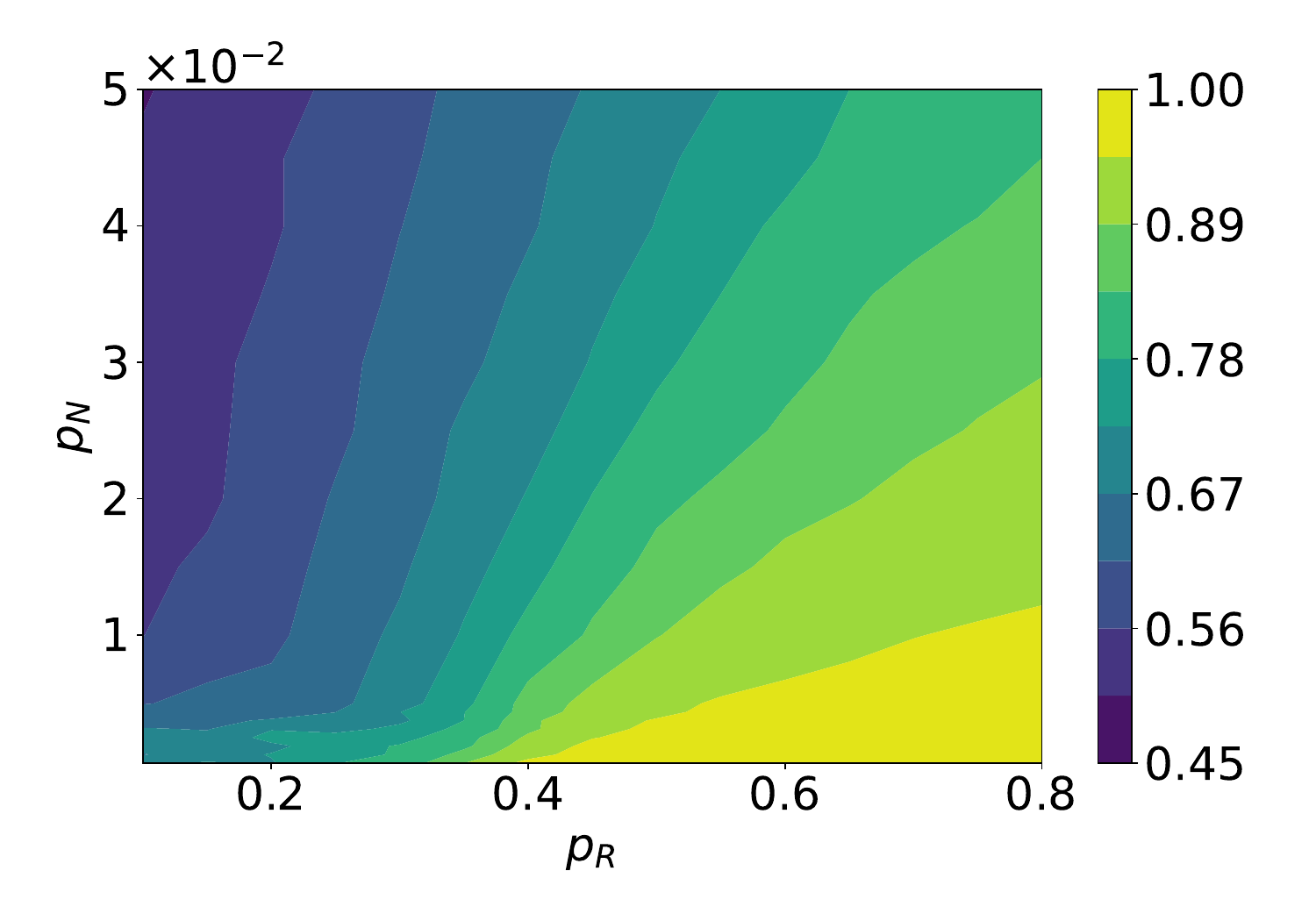}
    \captionsetup{justification=centering}
    \caption{Simulations with active rewiring, $q=4$ and $M=4$}
\end{subfigure}
\begin{subfigure}[t]{0.4\textwidth}
    \centering
    \includegraphics[width=\textwidth]{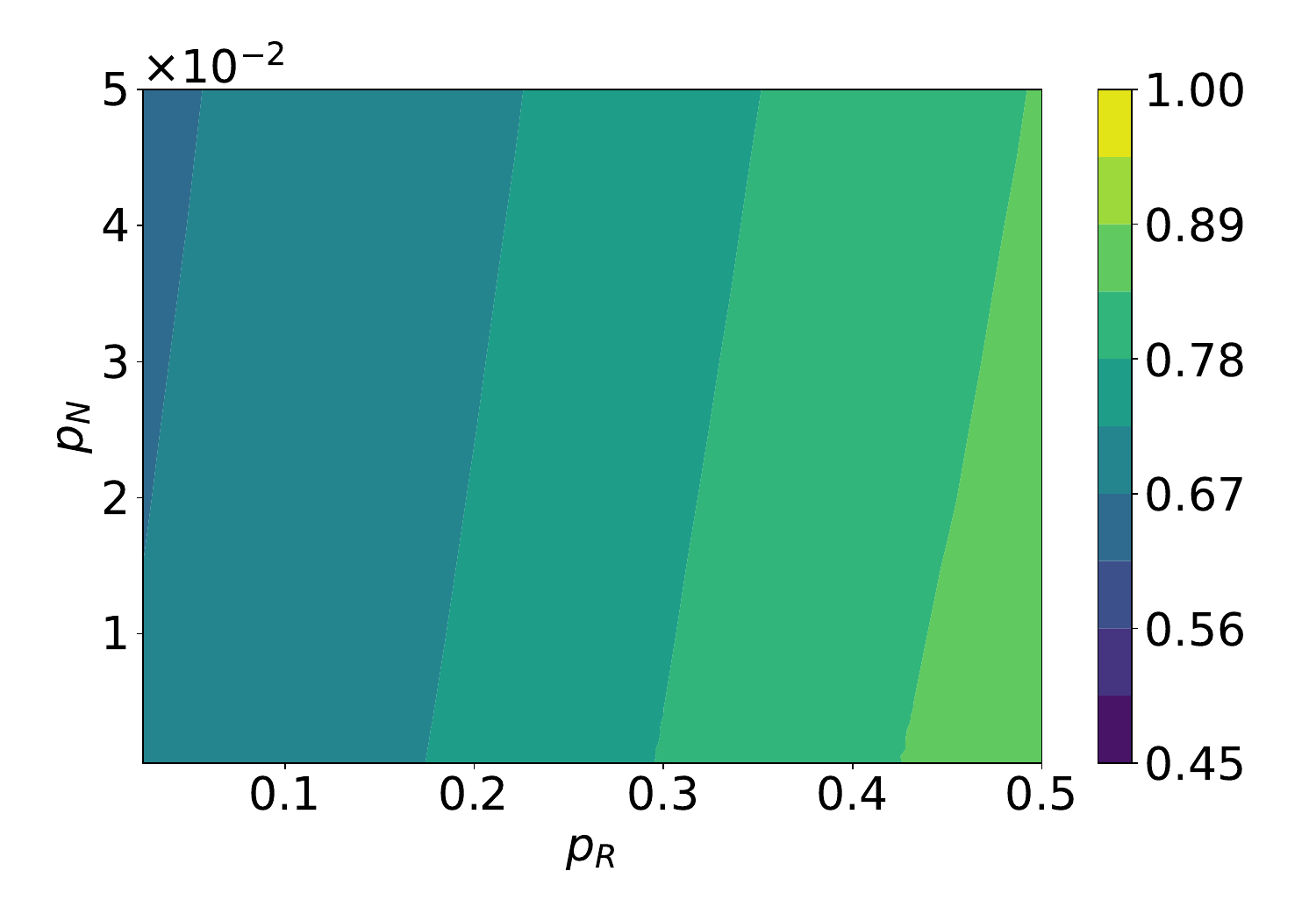}
    \captionsetup{justification=centering}
    \caption{Mean Field with reactive rewiring, $q=4$ and $M=2$}
\end{subfigure}
\begin{subfigure}[t]{0.4\textwidth}
    \centering
    \includegraphics[width=\textwidth]{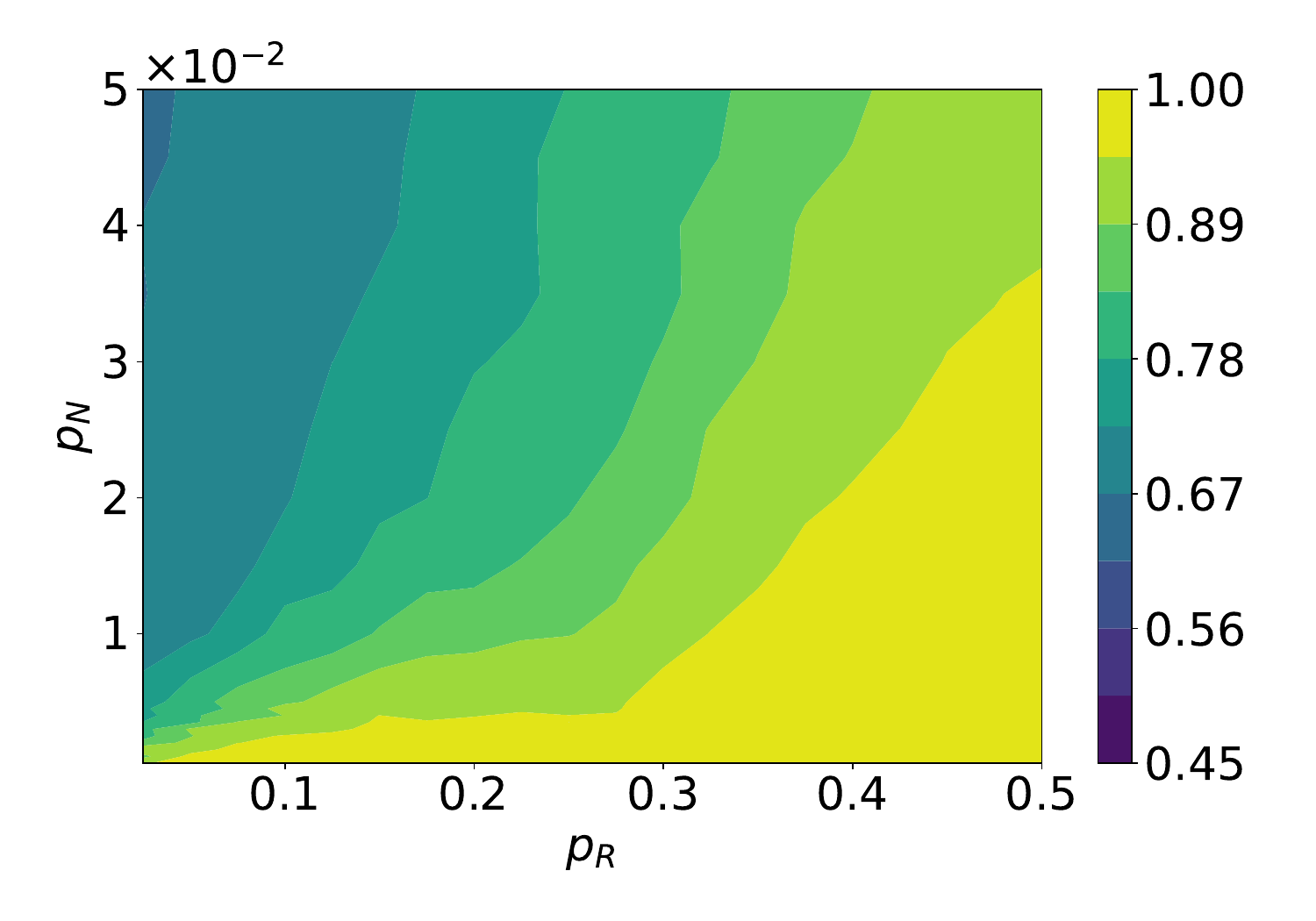}
    \captionsetup{justification=centering}
    \caption{Simulations with reactive rewiring, $q=4$ and $M=2$}
\end{subfigure}
\begin{subfigure}[t]{0.4\textwidth}
    \centering
    \includegraphics[width=\textwidth]{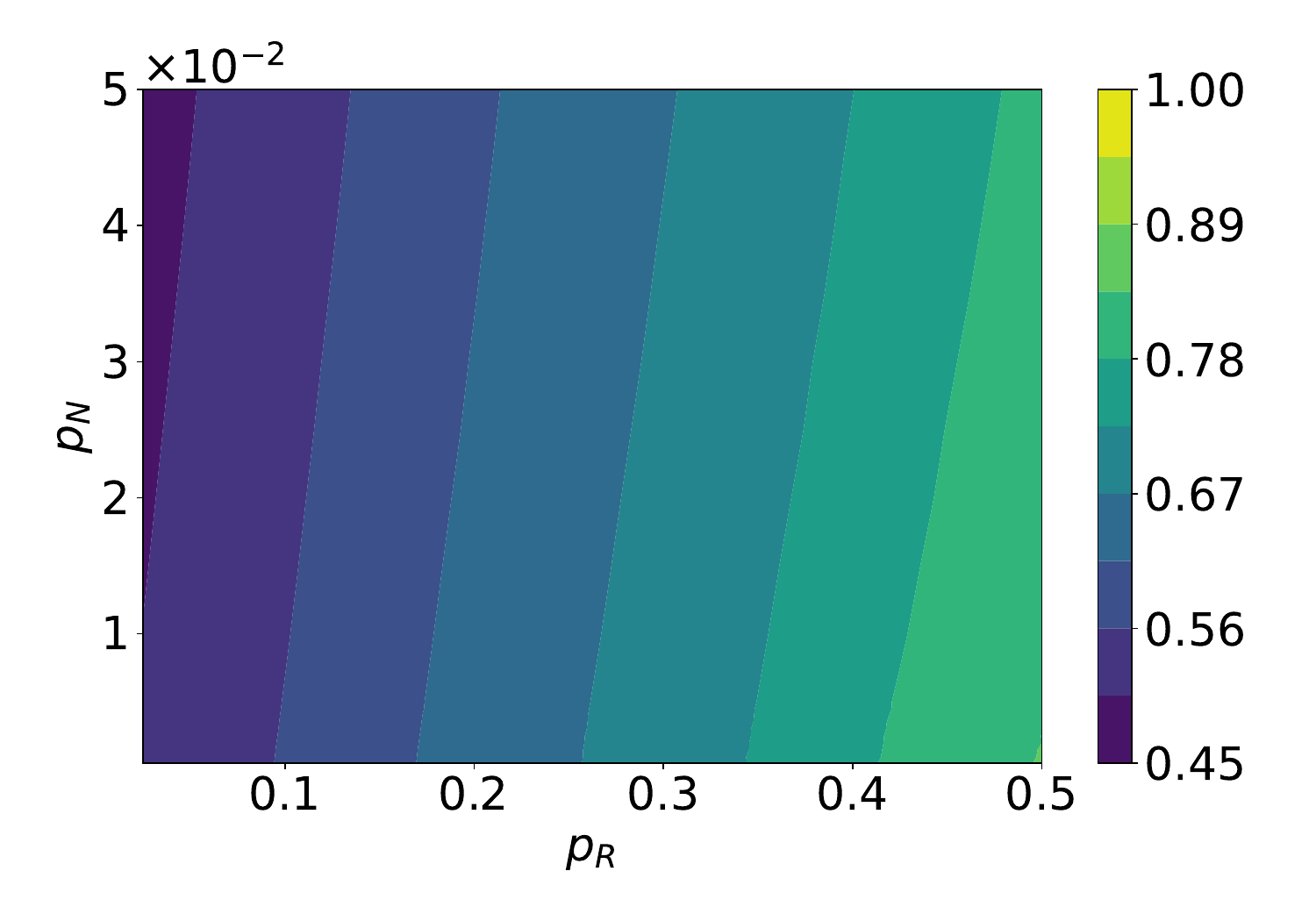}
    \captionsetup{justification=centering}
    \caption{Mean Field with reactive rewiring, $q=4$ and $M=4$}
\end{subfigure}
\begin{subfigure}[t]{0.4\textwidth}
    \centering
    \includegraphics[width=\textwidth]{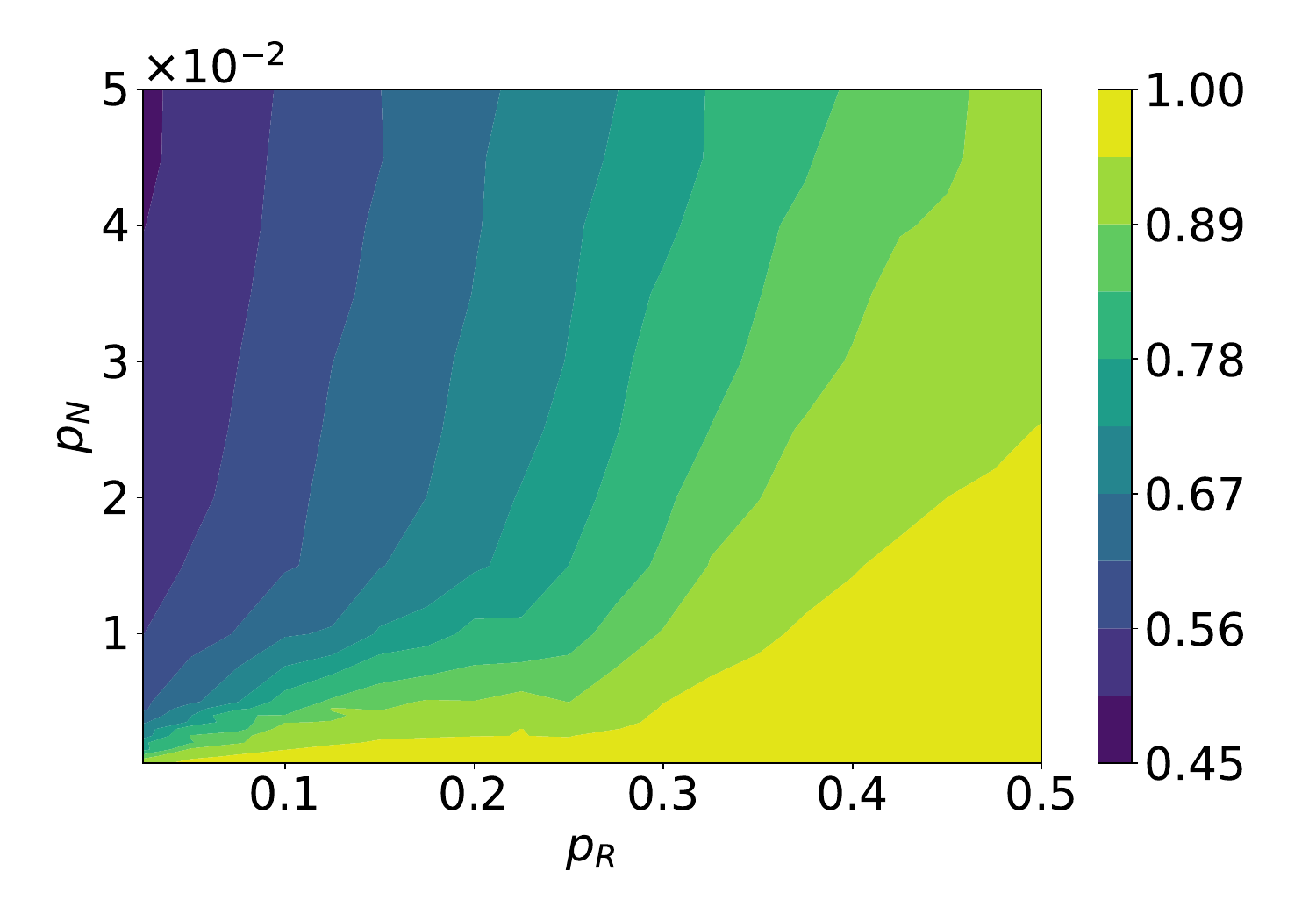}
    \captionsetup{justification=centering}
    \caption{Simulations with reactive rewiring, $q=4$ and $M=4$}
\end{subfigure}
\caption{Comparison between $S_{\infty}$ calculated by the mean field approximation (graphs to the left) and the value measured from the simulations (graphs to the right).}
\label{fig:MF-compare}
\end{figure}

A comparison between mean field and simulations can be found in figure \ref{fig:MF-compare}. We can see that the values predicted for the mean field approximation match the simulation results for higher values of $p_N$, while for low values discrepancies appear. These differences seem to be stronger for higher values of $M$ and in the reactive case. One possible explanation is that for lower values of $p_N$, even though every opinion holds the same number of sites when we consider averages over long times, most of the time is spent with an imbalance between the opinions, which violates the hypothesis of our mean field calculations.
\section{Conclusions}
\label{sec:conc}

In this work we studied the effects of the addition of noise in an adaptive voter model. Our main conclusion is that this change prevents the network where the model is being run from breaking into different components where each component holds only one opinion, as happens in the adaptive voter model without noise. Investigating the network structure reveals that what happens instead is that the network organizes itself into communities where there is a majority opinion. These communities can be identified using the stochastic block model.

\pagebreak

\onecolumngrid

\clearpage

\twocolumngrid

Investigating how intense this fragmentation is, some broad trends can be observed. We see that increasing the rewiring probability increases fragmentation and that increasing the noise intensity decreases fragmentation. While these are not surprising, the quantitative interplay between the two mechanisms is not entirely obvious. We also see that the number of opinions doesn't play a prominent role and that increasing $q$ prevents fragmentation in the active case, but promotes it in the reactive case.

In order to try to get some analytical results, a mean field treatment was developed, where a symmetry between all communities in the network is assumed. Comparing with the simulations we see that the analytical results roughly reproduce the qualitative behaviour of the simulations and for the case of high noise intensity they have a decent quantitative match.

\section{Acknowledgments}
The author gratefully acknowledges the Multiuser
Computational Center at UFABC (CCM-UFABC) for providing the
computational resources used in this study.

\onecolumngrid

\appendix

\section{Details of the Mean Field calculations}
\label{app:details}

In this appendix we will do the detailed derivation of equations (\ref{eq:contrib1}) to (\ref{eq:MF-reactive}). We recall that in our mean field treatment we make the following we make the approximation that the following holds at all times:
\begin{itemize}
\item All opinions are held by the same amount of agents at all times.
\item The probability $P_{\sigma, \sigma'}$ that an edge connects agents with opinions $\sigma$ and $\sigma'$ is
\begin{equation}
P_{\sigma, \sigma'} = 
\left\{
\begin{array}{ll}
p_{\mathrm{same}} & \mbox{ if }\sigma = \sigma' \\
p_{\mathrm{different}} & \mbox{ if }\sigma \neq \sigma'
\end{array}
\right.
\label{ap-eq:def-psame}
\end{equation}
With $p_{\mathrm{same}}$ and $p_{\mathrm{different}}$ independent of the specific opinions involved.
\item The degree distribution $P_q$ does not change over time.
\end{itemize}
We also recall that a timestep of the mean field model consists of: 

\begin{itemize}
\item Pick an agent $i$ at random.
\item With probability $p_N$, its opinion is changed at random (noise).
\item Otherwise, we draw the number of neighbours of $i$ (using the degree distribution) and how many of these neighbours agree with $i$ (using $p_{\mathrm{same}}$).
\item If $i$ is not isolated (that is, if the number of neighbours drawn in the last step is not 0), choose a neighbour $j$ and change $i$'s opinion to $j$'s opinion.
\item Do the appropriate rewires along the evolution (using $p_R$ for the probability).
\end{itemize}
Our objective is to find how a single timestep changes the probability $S$ that a pair of neighbouring agents have the same opinion, (note that $S = M p_{\mathrm{same}}$). Equations (\ref{eq:contrib1}), (\ref{eq:contrib2}) and (\ref{eq:contrib3}) give the contributions of different processes to the change in $S$ for active rewiring, while (\ref{eq:contrib4}) and (\ref{eq:contrib5}) give the contributions for reactive rewiring.

\subsection{Active rewires}

\subsubsection{First contribution}
The first contribution in equation (\ref{eq:contrib1}) covers the following situation:

\begin{itemize}
\item Agent $i$ was affected by the noise and its opinion was changed.
\item The total number of neighbours is $k$, drawn from $P_q$.
\item The number of neighbours holding the agent's previous opinion is $I\sim $ \,Binomial$(N=k,p=S)$.
\item The number of neighbours holding the agent's new opinion is $n\sim $ \,Binomial$(N=k-I,p=\nicefrac{1}{(M-1)})$.
\item The change in $S$ is entirely due to the change of the agent opinion and is given by $\nicefrac{(n - I)}{E}$, where $E = \nicefrac{Nq}{2}$ is the number of edges in the network.
\end{itemize}

Since the probability that this exact scenario plays out is $\frac{p_N(M-1)}{M}P(k, I, n)$ (affected by the noise and had its opinion changed, together with the drawn variables), then the average contribution to $\Delta S$ is given by the expectation over $k, I$ and $n$:

\[
\expect{\frac{p_N(M-1)}{M}\frac{(n - I)}{E}} = \frac{p_N(M-1)}{EM} \expect{n - I} = \frac{p_N(M-1)}{EM} \expect{\expectc{n - I}{I,k}} = 
\]\[
= \frac{p_N}{EM} \expect{\expectc{k-MI}{k}} = \frac{p_N}{EM} \expect{k-MkS} = \frac{p_Nq(1-MS)}{EM}
\]

\subsubsection{Second contribution}
The second contribution in equation (\ref{eq:contrib2}) covers the following situation:

\begin{itemize}
\item Agent $i$ was not affected by the noise.
\item The total number of neighbours $k$, drawn from $P_q$ is different from 0.
\item The number of neighbours holding the agent's current opinion is $I\sim $ \,Binomial$(N=k,p=S)$.
\item The neighbour chosen to interact with $i$ has a different opinion and their connection is rewired.
\item The change in $S$ is entirely due to the rewire and given by $\nicefrac{1}{E}$.
\end{itemize}

The probability that this exact scenario plays out is $p_N^{\complement}(1-\delta_{k,0})\frac{(k-I)}{k}p_R\,P(k, I)$ ($i$ is not affected by the noise, $k\neq 0$, the chosen neighbour has a different opinion and a rewire takes place, together with the drawn variables). So the average contribution to $\Delta S$ is given by the expectation over $k$ and $I$:

\[
\expect{\frac{\ppnc (1-\delta_{k,0})p_R (k-I)}{Ek}} = \frac{\ppnc p_R}{E} \expect{\expectc{\frac{(1-\delta_{k, 0})(k-I)}{k}}{k}} = \frac{\ppnc p_R}{E} \expect{\frac{(1-\delta_{k, 0})(k-kS)}{k}} = 
\]\[
= \frac{\ppnc p_R}{E} \expect{(1-\delta_{k, 0})(1-S)} = \frac{\ppnc p_R \ppoc \psc}{E}
\]

\subsubsection{Third contribution}
The third contribution in equation (\ref{eq:contrib3}) covers the following situation:

\begin{itemize}
\item Agent $i$ was not affected by the noise.
\item The total number of neighbours $k$, drawn from $P_q$ is different from 0.
\item The number of neighbours holding the agent's current opinion is $I\sim $ \,Binomial$(N=k,p=S)$.
\item The neighbour $j$, chosen to interact with $i$ has a different opinion and $i$ copies its opinion.
\item The number of neighbours (besides $j$) holding the agent's new opinion is $n\sim $ \,Binomial$(N=k-I-1,p=\nicefrac{1}{(M-1)})$.
\item The change in $S$ is entirely due to the change in $i$'s opinion and given by $\nicefrac{(n+1-I)}{E}$ (it now has $n+1$ neighbours agreeing with it, instead of $I$).
\end{itemize}

The probability that this exact scenario happens is $\ppnc (1-\delta_{k, 0})\frac{(k-I)}{k}\pprc P(k, I, n)$ ($i$ is not affected by the noise, $k\neq 0$, the chosen neighbour has a different opinion but a rewire does not happen). So we must compute the expectation:

\[
\expect{\ppnc (1-\delta_{k, 0})\frac{(k-I)}{k}\pprc\frac{(n+1-I)}{E}} = 
\frac{\ppnc\pprc}{E} \expect{\expectc{(1-\delta_{k, 0})\frac{(k-I)}{k}(n+1-I)}{k, I}} =
\]\[
= \frac{\ppnc\pprc}{E} \left(\expect{\expectc{(1-\delta_{k, 0})\frac{(k-I)}{k}(1-I)}{k}} + \expect{\expectc{(1-\delta_{k, 0})\frac{(k-I)}{k(M-1)}(1-I+k)}{k}}\right) =
\]\[
= \frac{\ppnc\pprc}{E} \left(\expect{(1-\delta_{k, 0})\left((S+1)\psc - kS\psc\right)} + \expect{(1-\delta_{k, 0})\frac{(k-1)\left(\psc\right)^2}{M-1}}\right) =
\]\[
= \frac{\ppnc\pprc}{E} \left(\ppoc(S+1)\psc - qS\psc + \frac{(q-\ppoc)\left(\psc\right)^2}{M-1}\right) = \frac{\ppnc\pprc\psc}{E(M-1)} \left(\ppoc(MS+M-2) + q(1-MS)\right)
\]

\subsection{Reactive rewires}

\subsubsection{First contribution}
The first contribution in equation (\ref{eq:contrib4}) covers the following situation:

\begin{itemize}
\item Agent $i$ was affected by the noise and its opinion was changed.
\item The total number of neighbours is $k$, drawn from $P_q$.
\item The number of neighbours holding the agent's previous opinion is $I\sim $ \,Binomial$(N=k,p=S)$.
\item The number of neighbours holding the agent's new opinion is $n\sim $ \,Binomial$(N=k-I,p=\nicefrac{1}{(M-1)})$.
\item Because of the opinion change, $r$ neighbours among the $k-n$ that disagree with $i$ rewire their connections; with $r\sim $ \,Binomial$(N=k-n,p=p_R)$.
\item The change in $S$ is due to the change of the agent opinion and the subsequent rewires, amounting to $\nicefrac{(n + r - I)}{E}$.
\end{itemize}

The probability of this scenario is $\frac{p_N(M-1)}{M}P(k, I, n, r)$ (affected by the noise and had its opinion changed). So we must obtain the expectation:

\[
\expect{\frac{p_N(M-1)}{M}\frac{(n + r - I)}{E}} = \frac{p_N(M-1)}{EM} \expect{\expectc{n + r - I}{k, I, n}} = \frac{p_N(M-1)}{EM} \expect{\expectc{n - I + p_R(k-n)}{k, I}} = 
\]\[
= \frac{p_N}{EM} \expect{\expectc{(k-I)\pprc + (k p_R - I)(M-1)}{k}} = \frac{p_N}{EM} \expect{k\left(M p_R - MS + Sp_R - 2p_R + 1\right)} = 
\]\[
= \frac{p_N q}{EM}(M p_R - MS + Sp_R - 2p_R + 1)
\]

\subsubsection{Second contribution}
The second contribution in equation (\ref{eq:contrib5}) covers the following situation:

\begin{itemize}
\item Agent $i$ was not affected by the noise.
\item The total number of neighbours $k$, drawn from $P_q$ is different from 0.
\item The number of neighbours holding the agent's current opinion is $I\sim $ \,Binomial$(N=k,p=S)$.
\item The neighbour $j$, chosen to interact with $i$ has a different opinion and $i$ copies its opinion.
\item The number of neighbours (besides $j$) holding the agent's new opinion is $n\sim $ \,Binomial$(N=k-I-1,p=\nicefrac{1}{(M-1)})$.
\item This change of opinion causes $r\sim $ \,Binomial$(N=k-n-1,p=p_R)$ rewires (note that $n+1$ neighbours agree with $i$'s new opinion because $j$ is not contabilized among the $n$).
\item The change in $S$ is due to the change of the agent opinion and the subsequent rewires, amounting to $\nicefrac{(n + 1 + r - I)}{E}$.
\end{itemize}

The probability of this scenario playing out is $\ppnc (1-\delta_{k,0})\frac{(k-I)}{k}P(k, I, n, r)$ (not affected by the noise, $k \neq 0$ and the neighbour has a different opinion). The contribution to $\Delta S$ is given by the expectation:

\[
\expect{\ppnc (1-\delta_{k,0})\frac{(k-I)}{k} \frac{(n + 1 + r - I)}{E}} = \frac{\ppnc}{E} \expect{\expectc{(1-\delta_{k,0})\frac{(k-I)}{k}(n + 1 + r - I)}{k, I, n}} = 
\]\[
= \frac{\ppnc}{E} \expect{\expectc{(1-\delta_{k,0})\frac{(k-I)}{k}\left((n + 1)\pprc + k p_R - I\right)}{k, I}} = 
\]\[
= \frac{\ppnc}{E} \expect{\expectc{(1-\delta_{k,0})\frac{(k-I)}{k}\left(k p_R - I + \pprc\left(1 + \frac{k - I - 1}{M - 1}\right)\right)}{k}} = 
\]\[
= \frac{\ppnc\psc}{E(M-1)}
\expect{(1-\delta_{k,0})\left((1 - k)\left(M\left(S + \pprc\right) - S p_R - 2 \pprc\right) + k(M-1)\right)}= 
\]\[
= \frac{\ppnc\ppsc}{E(M - 1)}\left(\left(M\left(\pprc +S\right) - 2\pprc - S p_R\right)\left(\ppoc - q\right) + q(M -1)\right)
\]

\twocolumngrid

\bibliographystyle{plain}
\bibliography{sociophysics}

\begin{thebibliography}{10}

\bibitem{rede-BA-def}
Albert~L\'aszl\'o Barab\'asi and R\'eka Albert.
\newblock Emergence of scaling in random networks.
\newblock {\em Science}, 286(5439):509--512, 1999.

\bibitem{echo-chamber-1}
Pablo Barberá, John~T. Jost, Jonathan Nagler, Joshua~A. Tucker, and Richard
  Bonneau.
\newblock Tweeting from left to right: Is online political communication more
  than an echo chamber?
\newblock {\em Psychological Science}, 26(10):1531--1542, 2015.

\bibitem{vot-majoritario-adapt}
E.~Burgos, Laura Hern\'andez, H.~Ceva, and R.~P.~J. Perazzo.
\newblock Entropic determination of the phase transition in a coevolving
  opinion-formation model.
\newblock {\em Phys. Rev. E}, 91:032808, Mar 2015.

\bibitem{vot-adapt-noise}
Philip~S. Chodrow and Peter~J. Mucha.
\newblock Local symmetry and global structure in adaptive voter models.
\newblock {\em SIAM Journal on Applied Mathematics}, 80(1):620--638, 2020.

\bibitem{echo-chamber-4}
Matteo Cinelli, Gianmarco De~Francisci Morales, Alessandro Galeazzi, Walter
  Quattrociocchi, and Michele Starnini.
\newblock The echo chamber effect on social media.
\newblock {\em Proceedings of the National Academy of Sciences},
  118(9):e2023301118, 2021.

\bibitem{vot-adapt-external-field}
Mario~G. Cosenza and José~L. Herrera-Diestra.
\newblock Coevolutionary dynamics with global fields.
\newblock {\em Entropy}, 24(9), 2022.

\bibitem{erdos-renyi}
Paul Erdos and Alfr\'ed R\'enyi.
\newblock On random graphs.
\newblock {\em Publicationes Mathematicae}, 6:290--297, 1959.

\bibitem{echo-chamber-2}
Matthew Gentzkow and Jesse~M Shapiro.
\newblock Ideological segregation online and offline.
\newblock Working Paper 15916, National Bureau of Economic Research, April
  2010.

\bibitem{livro-redes-adaptativas}
Thilo Gross and Hiroki Sayama.
\newblock {\em Adaptive Networks. Theory, Models and Applications}.
\newblock Springer Verlag, Germany, 2009.

\bibitem{votante-def}
R.~A. Holley and T.~M. Liggett.
\newblock Ergodic theorems for weakly interacting infinite systems and the
  voter model.
\newblock {\em Annals of Probability}, 3(4):643--663, 1975.

\bibitem{votante-spontaneous-rewire}
Petter Holme and M.~E.~J. Newman.
\newblock Nonequilibrium phase transition in the coevolution of networks and
  opinions.
\newblock {\em Phys. Rev. E}, 74:056108, Nov 2006.

\bibitem{vot-adapt-2}
Arkadiusz J\ifmmode~\mbox{\c{e}}\else \c{e}\fi{}drzejewski, Joanna Toruniewska,
  Krzysztof Suchecki, Oleg Zaikin, and Janusz~A. Ho\l{}yst.
\newblock Spontaneous symmetry breaking of active phase in coevolving nonlinear
  voter model.
\newblock {\em Phys. Rev. E}, 102:042313, Oct 2020.

\bibitem{vot-adapt-com-zealots}
Pascal~P. Klamser, Marc Wiedermann, Jonathan~F. Donges, and Reik~V. Donner.
\newblock Zealotry effects on opinion dynamics in the adaptive voter model.
\newblock {\em Phys. Rev. E}, 96:052315, Nov 2017.

\bibitem{vot-adapt-1}
Cecilia Nardini, Bal\'azs Kozma, and Alain Barrat.
\newblock Who's talking first? consensus or lack thereof in coevolving opinion
  formation models.
\newblock {\em Phys. Rev. Lett.}, 100:158701, Apr 2008.

\bibitem{graph-tool}
Tiago~P. Peixoto.
\newblock The graph-tool python library.
\newblock {\em figshare}, 2014.

\bibitem{echo-chamber-3}
H.~Akin Unver.
\newblock Digital challenges to democracy: Politics of automation, attention,
  and engagement.
\newblock {\em Journal of International Affairs}, 71(1):127--146, 2017.

\bibitem{deffuant-adapt-noise}
Y.~Yu, G.~Xiao, G.~Li, W.~P. Tay, and H.~F. Teoh.
\newblock {Opinion diversity and community formation in adaptive networks}.
\newblock {\em Chaos: An Interdisciplinary Journal of Nonlinear Science},
  27(10):103115, 10 2017.

\end{thebibliography}

\end{document}